\documentstyle[psfig,12pt]{article}
\begin{document}
\baselineskip = 14pt
\addtolength{\baselineskip}{\baselineskip}

\title{
Microscopic Motion of Particles \\
Flowing through a Porous Medium}

\author{Jysoo Lee$^{(1,2)}$ and Joel Koplik$^{(2)}$ \\ 
\\
$^{(1)}$ Department of Physics, Seoul National University \\
Seoul 151-742, Korea \\
\\
$^{(2)}$ Benjamin Levich Institute and Department of Physics \\
City College of the City University of New York \\
New York, NY 10031}

\date{\today}

\maketitle

\vspace{0.5in}

\begin{abstract}

We use Stokesian Dynamics simulations to study the microscopic motion
of particles suspended in fluids passing through porous media.  We
construct model porous media with fixed spherical particles, and allow
mobile ones to move through this fixed bed under the action of an
ambient velocity field.  We first consider the pore scale motion of
individual suspended particles at pore junctions.  The relative
particle flux into different possible directions exiting from a single
pore, for two and three dimensional model porous media is found to
approximately equal the corresponding fractional channel width or
area.  Next we consider the waiting time distribution for particles
which are delayed in a junction, due to a stagnation point caused by a
flow bifurcation.  The waiting times are found to be controlled by
two-particle interactions, and the distributions take the same form in
model porous media as in two-particle systems.  A simple theoretical
estimate of the waiting time is consistent with the simulations.  We
also find that perturbing such a slow-moving particle by another
nearby one leads to rather complicated behavior.  We
study the stability of geometrically trapped particles.  For simple
model traps, we find that particles passing nearby can ``relaunch''
the trapped particle through its hydrodynamic interaction, although
the conditions for relaunching depend sensitively on the details of
the trap and its surroundings.

\end{abstract}

\newpage
\baselineskip = 14pt

\section{Introduction}

The transport of particulate suspensions through porous media is a
process with numerous industrial applications such as deep bed
filtration \cite{hll70,g77,dbl88,t89}, hydrodynamic chromatography
\cite{chrom}, migration of fines \cite{kfa83}, ground water
contamination \cite{ground}, the flow of dilute stable emulsions
\cite{ssr84}, and hindered diffusion in membranes \cite{sj89}.  In
order to understand the behavior of these systems, one often needs to
know the microscopic, or ``pore-scale'' behavior of the suspended
particles.  Consider the example of deep bed filtration, where a
suspension is injected into a filter made of porous material, in the
hope that the suspended particles will be collected in the filter
while clear(er) fluid passes through.  A filtrate particle
flowing through the pore space may be trapped by the geometric
constraint of reaching a pore smaller than its diameter, or by other
adhesive mechanisms such as electrostatic or van der Walls forces.
Realistic porous media have an intricate randomly-sized and
randomly-interconnected pore space with highly nontrivial flow paths.
The degree to which suspended particles choose trajectories in the
pore space with or without small constrictions, or tend to be
attracted or repelled by the walls, or move towards or away from the
other suspended particles will determine the dynamics of the
filtration process.  Recent experiments \cite{gdhg96a,gdhg96b} which
focus on such microscopic particle behavior in filtration clearly
illuminate how such pore-scale details can change the macroscopic
properties of the system.

A useful way of describing the porous medium and particulate motion
therein is to note that the pore space of a material like a random
sphere pack involves relatively large open regions connected by
relatively narrow channels: pores and throats, respectively
\cite{dul92}.  A typical pore is connected to several others through
the throats.  The streamlines for fluid flow or the paths taken by
suspended particles (in the absence of wall adhesion mechanisms) are
then roughly unidirectional within the throats but branch at each
pore, corresponding to the different connected neighboring pores which
are accessible.  A key issue for microscopic particulate motion is the
``junction rule'': when a suspended particle reaches a pore, which of
the neighboring pores does it choose to move towards next.

The necessity for such information is evident when constructing
a quantitative description for the system.  Consider, for example, the
network model for filtration, which models a filter medium as a
ball-and-stick network of nodes interconnected by channels
\cite{lmdg84,rf88,si91,hsd93}.  The trajectory of the suspended
particles in the network is largely determined by the motion of the
particles at the nodes, which is specified by the junction rule.
Furthermore, the interaction between two nearby particles has been
experimentally found to be important \cite{gdhg96a,gdhg96b}.  For
example, a particle which is apparently trapped at some point in the
pore space may escape from the trap if another suspended particle
passes nearby. This effect can significantly change the distribution
of the trapped particles, and such information must be included in the
model.  The problem which motivates this work is that very little
quantitative information on these crucial effects is available.  For
the junction rule, there have been studies on the related problem of
red cells passing through tube hematocrits
\cite{j71,gpa79,ddc83,bh64,ao87}, where similar but different
geometries were considered.  In the filtration case, only rough
estimates exist for both the junction probabilities and for the escape
rate of trapped particles \cite{gdhg96a,gdhg96b}.

In this paper, we study the microscopic motion of particles suspended
in a fluid passing through a porous medium, using Stokesian Dynamics
(SD) computer simulations \cite{bb88}.  We construct our model porous
media by fixing the positions of a set of spheres, representing the
grains of the porous medium, and allowing other particles,
representing the filtrate, to move through this fixed bed.  The
original SD code, which simulated only mobile particles, was modified
appropriately to handle fixed particles as well.  Given that we are
interested here in relatively large non-colloidal particles, whose
size is in the micron range, Brownian forces are not included in the
calculations.  Using this methodology, we describe the filtrate motion
at junctions in two and three dimensional model porous media in terms
of a quantity called the fractional particle flux (FPF).  At a given
pore, the FPF for any subsequent connected channel is defined as the
fraction of particles choosing it out of all possible exit
channels.  We find that it is a good approximation to assume that
the FPF of a channel is proportional to its cross-sectional area.
Although this result agrees with one's simplest intuitive expectation,
we are not aware of any previous quantitative verification.

A suspended particle may spend a relatively long time in a pore
before proceeding to an exit channel.  To quantify this effect we have
also measured the ``waiting'' time for a particle passing through two
and three dimensional model porous medium, and also for a mobile
particle passing near one fixed particle.  These calculations show
that the waiting time is dominated by the two particle interaction --
that between the mobile particle and fixed particle which divides the
two exit channels.  We also present an approximate analytic
calculation for the waiting time, which is consistent with the
numerical results.  We further study the effect of another mobile
particle on ``hesitating'' particles -- those near bifurcating streamlines
with large waiting times.
The resulting motion of the two particles displays an unexpectedly
rich range of possibilities, such as ``push toward'', ``push away'',
``turn'', and ``lead'', which we describe in more detail below.

Another important result is the condition under which geometrically
trapped particles escape from the trap by the perturbation of a
``bypassing'' particle -- another suspended particle passing nearby.
We construct simple traps consisting of two and three fixed particles
in two and three dimensions, respectively, and place a mobile particle
in the trap.  In agreement with observations, We find that a bypassing
particle can ``relaunch'' the trapped particle based upon only
hydrodynamic interactions.  We determine the trajectories of the
mobile particles which cause relaunching for the parameters
characterizing the trap.  We find that the condition for the
relaunching depends not only on the geometry of the trap, but also on
the surroundings.  We then study the condition for the relaunching in
a two dimensional model porous medium, and obtain qualitatively
similar results.

\section{Stokesian Dynamics}

Stokesian Dynamics (SD) is a numerical method to dynamically simulate
the behavior of solid particles suspended in a viscous fluid at zero
Reynolds number \cite{bb88}, based on the
resistance matrix description of particle motion in creeping flow
\cite{bo72}. When the Reynolds number based on the particle length
scale is negligible, the generalized hydrodynamic forces $\vec{F}^{H}$ 
exerted on the particles in a suspension are
\begin{equation}
\label{eq:resist}
\vec{F}^{H} = - \mu ~ R_{FU} \cdot (\vec{U} -\vec{U}^{\infty}),
\end{equation}
where the components of $\vec{U}$ are the generalized particle velocities,
$\vec{U}^{\infty}$ is the ambient flow velocity evaluated at the
particle centers, and we have assumed that there is no bulk shear
flow.  The term generalized means that both translational and rotational
components are included:  the
velocity vector contains angular velocity components, and torques are
included in the force vector, so that $\vec{U}$ and $\vec{F}^{H}$ have
$6N$ components, where $N$ is the total number of the particles.
$R_{FU}$ is the configuration dependent resistance matrix that gives
the hydrodynamic forces on the particles due to their motion relative
to the fluid.  The inverse of the resistance matrix is the mobility
matrix $M_{FU}$.

We briefly describe the calculation of $R_{FU}$ in the SD method; a
detailed discussion can be found in \cite{bb88}.  The force density on
the surface of each particle is expanded in a series of multipole
moments, where terms higher than quadrupole are ignored.  The far
field (large interparticle distance) value of the mobility matrix
$M_{FU}$ is expressed in terms of these moments.  We invert $M_{FU}$
to get the form of the resistance matrix $R_{FU}$ relevant to the far
field region.  We then add near field (short interparticle distance)
lubrication terms to the matrix.  The resulting matrix $R_{FU}$ is
accurate both for far and near fields, but not necessarily for
intermediate distance.  In order to determine accurate values of
$R_{FU}$ at intermediate distance, Stokes equation is solved
numerically for the system of two spheres with various separations.
The numerical values of $R_{FU}$ from the simulations are included in
the present code.

Neglecting inertia, the total force on each particle is zero,
\begin{equation}
\label{eq:balance}
\vec{F}^{H} + \vec{F}^{P} = 0, 
\end{equation}
where $\vec{F}^{P}$ are the non-hydrodynamic forces acting on the
particles, which we specify presently.  Combining Eqs. (\ref{eq:resist}) 
and (\ref{eq:balance}),
one can calculate the velocities of the particles, and from the
velocities, one can calculate the positions of the particles at a
short time later.  For this new configuration, the velocities can
again be calculated, from which a new configuration can again be
generated, and this process can be repeated to generate the trajectories
of the particles.  In this paper we wish to apply this method to the
study the flow of suspensions through small subregions of rigid porous
media.  We construct a model porous medium by fixing the positions of
certain particles to be at chosen locations, which requires some modifications
of the SD method.  The basic equation (\ref{eq:resist}), the
integration method for the (force free) moving particles, and the resistance
matrix remain unchanged.  However, a further requirement $\vec{U} = 0$ is
required for the fixed particles, and in effect we will choose the
value of an appropriate non-hydrodynamic force to impose this
condition.

In the original SD formulation the velocities of the particles are
determined by calculating $\vec{F}^{H}$ from the force-free condition
Eq. (\ref{eq:balance}), and inverting Eq. (\ref{eq:resist}).  To fix
particle's position, we apply an additional ``gluing force'' to
balance hydrodynamic and other non-hydrodynamic forces.  The problem
is that the value of the gluing force is given only as a part of the
solution of the problem, and we cannot simply invert
Eq. (\ref{eq:resist}) to obtain the velocities.  We instead calculate
the velocities by a recursive method.  For given values of the gluing
forces, one can calculate the velocities of all the particles by
inverting Eq. (\ref{eq:resist}).  The resulting velocities are, in
general, not the correct solution of the problem, that is, the
velocities of the supposedly fixed particles are not equal to zero.
We define $\Sigma$ as the sum of the squares of the speed of the fixed
particles, which is then a function of the gluing forces.  The correct
solution is obtained for the values of the gluing forces which makes
$\Sigma = 0$, and since it is a non-negative quantity, the problem is
to find the global minimum of $\Sigma$ in the parameter space of the
gluing forces.  In other words, the problem becomes the minimization
of a function, and the method of the steepest descent \cite{ptvf92} is
effective.  In practice, we begin the iterative solution by using the
gluing force from the previous time step, and deem that a solution is
obtained when the average speed of the fixed particles is less than
$10^{-5}$ of the ambient velocity.  Fortunately, it turns out that
$\Sigma$ is a smooth function of the gluing forces, and does not
contain other local minima, and typically $50$ iterations suffice at
each timestep.

To check the code, we examined a few simple two-particle configurations
where an analytic solution is available.  At present, only
three-dimensional SD codes for monodisperse particles exist,
and in the remainder of this paper, the radius of all particles is set 
to $1$.  First, we consider
head-on collisions, where we fix a particle at $(0,0,0)$ and place a
moving particle at $(-r,0,0)$ in an ambient velocity field
$\vec{U}^{\infty} = (1,0,0)$.  (In all cases considered in this paper, 
no ambient angular velocity is applied, and since we are interested in the
trajectories of spheres, for the most part we suppress the discussion of 
their individual angular velocities.)  Our numerical results for the
velocities are compared with those obtained analytically using the
expression for the resistance matrix $R_{FU}$ given in \cite{kk91}.
The two velocities agree to within $10^{-3}$ percent for both near and
far fields.  We then consider tangential motion by placing the moving
particle initially at $(0,r,0)$, without changing the other 
parameters.  The quality of the agreement is similar to the head-on
collision case.  Another check is whether the solutions from the
simulations exhibit required symmetries.  We consider placing the
moving particle initially at $(\pm x_0, y_0, 0)$, with the fixed
particle and ambient flow field as above.  From the obvious
geometrical symmetries of this system, the $x$ velocity, the absolute
values of the $y$ velocity and the $z$ angular velocity should be the
same, but the signs of the $y$ velocity and the $z$ angular velocity
should be reversed for the two settings.  For several combinations of
$x_0$ and $y_0$, we confirm that the solution indeed possesses the
required symmetry.  Lastly, we also check the dependence of the
solution on the size of the computational domain.  A periodic boundary
condition is applied in each direction.  We find that the linear size
of $10,000$ is enough to ensure that the effect of the boundary is
negligible.

Particles in an inertialess suspension cannot overlap with or even
touch each other, since the radial component of the lubrication force
diverges when two particles approach.  In the simulation, however, we
find that the particles do overlap at times, due to the following
numerical subtlety. When a moving particle approaches a fixed one, at
sufficiently small interparticle distances the lubrication force
indeed becomes large, and it pushes the particle away.  However, since
the force grows slowly, the distance at which the repulsion occurs
often is very small.  In our simulation, we find that this distance is 
typically smaller than $10^{-13}$, and in principle one has to calculate the
trajectory of the particles to extremely high precision.  The problem
is that the simulation time with this high accuracy is enormous:
double precision computations are inadequate, and 
quadruple precision variables are needed, and furthermore the time step 
has to be
kept very small: less than $10^{-4}$.  The CPU time required for even a two
particle simulation at this accuracy exceeds 50 hours on 
a DEC AlphaStation 500/500, so such an SD simulation in a model porous 
medium is not feasible.

This type of simulation is not only practically impossible, but also
not at all physical.  The surface roughness of ordinary solid
particles is typically $10^{-3}$ to $10^{-2}$ of the radius
\cite{sl89}.  As a result the particles are not exactly spherical, and
their flow behavior changes significantly when the two particle
separation is at the scale of the roughness \cite{d92}.  One
semi-physical way to include the effects of roughness is to add an
extra repulsive force at very short distances \cite{ds96}.  In this
paper, we instead follow the simpler procedure of Phung and Brady
\cite{p93}, who on one hand introduced a cutoff in the resistance
matrix computation, and on the other simply ignore small overlaps.
When interparticle gap is smaller than the cutoff value of $10^{-5}$,
the value of $R_{FU}$ at the cutoff is used.  Accidental overlaps are
allowed, provided the overlap distance is smaller than $10^{-3}$.

\section{Fractional Particle Flux}

Most pores in typical porous media have more than one exit channel
connected to them, and the path taken by a particle in such a pore in
proceeding to the next one depends in detail on microscopic variables
such as the detailed local geometry, the particle's precise location,
and the flow field (which in turn depends on the location of the other
particles).  It is not feasible to study more than a small subregion
of a porous medium in such fine detail, and it is useful to construct
a more tractable model based on a network of nodes and links.  A key
ingredient is such a coarse-grained description is the fraction of
particles passing through a given exit path, averaged over some of the
microscopic variables.  We refer to this quantity as the fractional
particle flux (FPF).

Although the FPF is important in understanding porous media transport
in general and processes such as filtration in particular, we are not
aware of any direct measurement of this quantity.  Although it is
possible to follow the motion of a suitably tagged suspended particle
in some detail through a laboratory porous medium, one would have to
map out the microscopic pore space as well, and the practical
difficulties are evident.  A related problem which has received some
quantitative study is the flow of red cells passing through tube
hematocrits.  When a tube hematocrit bifurcates to two or more smaller
tubes, a blood cell has to choose a tube at the junction.  The
fractional red cell flux is of great importance in determining red
cell distribution among the microvessels, and there exist several
direct and indirect measurements of it \cite{j71,gpa79,ddc83,bh64,ao87}.
The essential problem in these two systems is the same, but their
geometries are a bit different and blood cells are deformable, which
may result in different qualitative behaviors of the FPF.

We construct model porous media with fixed particles in two and three
dimensions.  First, we consider a two dimensional medium, where
particles are confined to a plane, immersed in a three dimensional
fluid.  Although a two dimensional medium is not realistic, it allows
us to study more detailed properties (note that computation time in SD
simulations increases roughly as the square of the number of
particles).  Fortunately, we shall see that its qualitative behavior
does not differ from that of a three dimensional medium, which we discuss
later.  The medium consists of the $11$ fixed particles shown in Fig.~1(a).
The centers of all particles are in the $z = 0$ plane, and we choose
the coordinates of the lower left particle to be the origin.  The
``lattice constant'' $a$ is the distance between neighboring particles
in the same column, and it also is the distance between neighboring
columns.  The middle column can be shifted vertically, and we choose
the $y$ coordinate of the center particle to be $y_{c} + b$, where
$y_{c} = 3a/2$.  We call the system, characterized by two parameters
$a$ and $b$, as the ``2d-11'' geometry.  Recall that the current
code can handle only monodisperse particles whose radius is taken as
$1$.  The ambient flow field is $(1,0,0)$ in all cases, unless stated
otherwise.  We insert a moving particle at $(-2,y_c+d,0)$, with
$-a/2< d < a/2$, just upstream
of the first column (Fig.~1(b)).  After the particle reaches the
node behind the first column, it will proceed to either channel A or B
depending on the initial parameter $d$.

Assuming that the distribution of the particle along the starting line
($x = -2$) is uniform \cite{note}, we can calculate the FPF from the
range of $d$ within which the trajectory of the particle passes
through channel A (or B).  The FPF for a channel is often plotted
against the fractional flow rate through the channel, but in the
present simulation the flow rate of a channel is difficult to
calculate.  We use the fractional channel width (FCW) instead.  
We define $W_{i}$ as the width of the $y$ interval corresponding to channel 
$i$ within the unit cell $[y_{c}-a/2,y_{c}+a/2]$, as shown in Fig.~1(b).
The FCW of channel $i$ is defined as $W_{i}$ normalized by the lattice
constant $a$.

We consider three values of the lattice constant, $a=4.25$, $4.5$, and
$5.0$, and we also vary the fractional channel width by changing $b$,
using five values of $b$ for each $a$.  
For a given geometry, determined by $a$ and $b$, we can measure the
FPF of a channel by studying the trajectories of the particles
starting from different locations characterized by $d$.  Note that
we can determine the FPF with less effort by focusing on a neighborhood of
the value
of $d$ at which the trajectory of the particle terminates at the
center particle.
The calculation of a typical trajectory requires about $4$ CPU hours on
a workstation,  About 15 trajectories are needed to determine
the FPF for a given geometry with fair accuracy.  The FPF for the
channel A are shown against its FCW in Fig.~2.  In this geometry, the
FPF for FCW less than $1/2$ can be determined simply by symmetry.

The most prominent feature of the figure is that all the curves lie
very close to the line:  FPF = FCW.  In other words, it is a good
approximation to distribute particles in proportion to the exit channel
width.  Small, but finite deviations from the curve are observed,
especially for large FCW.  Somewhat similar behavior is observed in
simulations of the flow of red blood cells at a capillary bifurcation
\cite{ao87}, but there are two noticeable differences.  In the blood
cell case, the FPF is larger than FCW for large FCW, in contrast to
the present case.  By inspecting the trajectories, it seems that the
difference is caused by the difference in the local geometry around
the junction.
Another difference is that in the present case, the amount of the
deviation increases as $a$ increases, in contrast to the cell case.
This is a bit puzzling, since we expect the FPF to follow the relative
width for larger lattice constants.  Further simulations indicate that 
the deviation shows a complicated dependence on $a$, before beginning to 
decrease for larger values, $a > 10$.

We next study a model three dimensional porous medium, consisting
of 13 fixed particles, in three layers, forming a deformable hexagonal
close packed (hcp) structure.  There are first and third layers
containing three particles each, and a second (middle) layer with
seven particles, as shown in Fig.~3(a), where the layers are 
stacked in the $x$-direction normal to the plane of the figure.  The
$y$ and $z$ coordinates of the particles in the first and third layers
are in register, while the middle layer can be translated in its plane to
produce a family of porous structures.  Specifically,
the $y$ coordinate of the center particle in the
second layer lies in the dotted (center-)line of the first layer, and
is shifted in the $z$ direction by a variable amount $b$.  When $b=0$ 
the orthocenter (small circle) in the first layer coincides with the
center particle in the second layer, and the particles form precisely 
a hexagonal close packed cell.  The second parameter describing
the geometry is the distance between nearest neighbor particles in the same
layer, the lattice constant $a$. The interlayer distance is fixed at
$a(2/3)^{1/2}$ in all cases studied.  This system is referred to below 
as the ``3d-13'' geometry.

The ambient flow again has unit strength in the $x$ direction normal
to the layers.  We insert a mobile particle in front of the gap in the
first layer and, depending on its initial coordinates, it will
eventually be carried around the central particle in the second layer
and proceed to one of the exit channels A, B or C shown in Fig.~3(b)
The initial $y$ and $z$ coordinates of the mobile particle lie within
the ``unit cell''---the largest triangle in Fig.~3(b), where the
projection of the particles to the $y-z$ plane is shown.
Again, the question is the fractional particle flux through each
channel.  Due to symmetry, the FPFs of channels A and B are
identical, and all three fluxes add to unity, so it suffices to give
the FPF of channel C.  
As in the previous simulation, calculating the FPF involves finding
the ranges of initial coordinates such which the trajectory of the
mobile particle passes through channel C.  However, a rather larger
number of trajectories for a given cell, 30, are required here, so we
consider only the case $a = 10$, and the four values $b = -2, 0, 2,
4$.  The calculation of one trajectory now requires about $6$ CPU
hours.

The measured FPF for channel C is plotted against its fractional
channel area (FCA) in Fig.~4.  
The FCA of channel $i$, which is an analog of the FCW in three
dimensions, is defined as the projected area of the triangle $W_{i}$,
normalized by the area of the unit cell---$\sqrt{3}a^{2}/2$.
The FPF curve lies again very close to the FPF = FCA line, and the
number of particles passing through a channel is to a very good
approximation proportional to its channel area.  Small deviations from
the FPF = FCA line can be seen, similar to those in the 2d-11
geometry.

From these simulations of motion in two and three dimensional porous
media, we find that the fractional particle flux for a channel is, to
a good approximation, proportional to its channel width (in 2d) or area
(in 3d).
Since a particle at a given initial location always passes through the
same channel, we expect the ``mixing'' at the node is not significant,
which is consistent with observations in the flow of red blood
cells \cite{ao87}.  A contrasting assumption is commonly made for the
motion of passive tracers in porous media flow.  The alternative
``complete mixing'' rule assumes that a tracer particle (effectively,
a particle small enough not to disturb the flow field) in a pore
chooses an exit pore based only on the relative flux there,
independent of its position of entry into the pore.  Evidently, this
assumption is reasonable only if the local P\'eclet number is small
enough for the particle to diffuse substantially within the pore and
lose memory of its initial streamline.

\section{Waiting Time near a Junction}

Consider a particle in a pore with, for example, two possible exit
paths, In the absence of suspended particles, the flow field will have
a dividing streamline terminating on the fixed particle, and a passive
fluid particle on this streamline would reach the fixed particle only
after infinite time.  Suspended particles will alter this picture
somewhat, but there will be a single mathematical trajectory which
reaches the fixed particle after infinite time, and when a particle
lies close to the line its velocity will be small, and it will
``wait'' in this pore before proceeding to the next.  Such behavior
was observed in the simulations of a particle passing through a
bifurcating tube \cite{ao87}.  Here, we study in detail the parameter
dependence of the waiting time for model porous media.

We first study the waiting time numerically for the 2d-11 geometry.  A
mobile particle is inserted at $(-2, y_c + d)$, just upstream of the
first layer (see Fig.~1(b)).  This particle may dally in front of the
center particle which divides channels A and B, and to quantify this
effect we define the waiting time $T_w$ as the time the mobile
particle is in the interval $(x_c - 2) - 0.1 < x < (x_c - 2) + 0.1$, where
$x$ ($x_c$) is the $x$ coordinate of the mobile (center) particle.  In
Fig.~5(a), we plot $T_w$ against $d$ for $a = 4.5$ and five values of
$b$.  (For other values of $a$, 4.25 and 5.0, the qualitative features
of the results are unchanged.)  As seen in the figure, $T_w$ is
sharply peaked around a $b$-dependent value $d_{\rm peak}$, where
waiting times as much as 100 times larger that the large-$d$ values
are seen.  Furthermore, the peaks are essentially all the same, as
seen if we superpose the individual peaks by plotting $T_w$ against $d
- d_{\rm peak}$ in Fig.~5(b), and noting that they roughly collapse to
a single curve.  Since the individual peaks intuitively correspond to
the moving particle stagnating near the dividing fixed particle in the
second layer, and do not seem to depend on exactly where the latter is
located, one suspects that the waiting time is only sensitive to the
two particle interaction between the mobile and the center particles.

Similar behavior is found in three dimensions.  In the 3d-13 geometry,
we place a mobile particle in front of the first layer, along the
vertical line passing through the center particle of the second layer
in Fig.~3(b).  Here, we define $d$ as the difference in $z$ coordinate
between the mobile particle and the center particle of the second
layer, and the waiting time $T_w$ as the amount of time the mobile
particle spends in $x_c - 2 - 0.1 < x < x_c - 2 + 0.1$, where $x$
($x_c$) is the coordinate of the mobile (center) particle.  In
Fig.~6(a), we show $T_w$ for $a = 10$, and four values of $b$.  These
curves are similar to those for the 2d-11 geometry and, when shifted
by their peak positions $d_{\rm peak}$, collapse again to the single
curve in Fig.~6(b).  The collapsed curve is very similar to that of
the 2d-11 geometry, further supporting the idea that the effect of the
particles surrounding the mobile and center particles is not
significant.

To pursue this simplifying idea, we consider the interaction of a
mobile particle with only a single fixed particle.  We fix the latter
at the origin, and place a mobile particle initially at $(-5,d,0)$.
We again define the waiting time $T_w$ as the time the mobile particle
spends in $- 2.1 < x < - 1.9$.  In Fig.~7, we plot the waiting time
for the two particle system in a log-log plot, along with those of the
2d-11 ($a = 4.5, b = 0$ and $b = 0.45$) and the 3d-13 geometries ($a =
10, b = 0$).  Aside from a difference in overall scale, the waiting
time variation for the different cases is identical, verifying that
the two-particle interaction is the dominant factor.  (The origin of
the difference in time scale is simply a matter of a different
superficial velocity: the asymptotic velocity far from the packing is
the same in all cases, so the scale of the velocity near the two
particle subsystem will depend on the superficial velocity within the
packing.  The 2d-11 geometry is most closely packed, has the lowest
permeability and lowest velocities, and the longest times.  The two
particle system is the most open and has the highest velocities and
shortest times.)

The two particle system can be treated analytically.
We fix a particle at the origin and insert a mobile particle at
$(-2-d,0,0)$, where $0 < d \ll 1$ for the long waiting time limit.
The ambient velocity field is $U^{\infty} =
(\cos\theta,\sin\theta,0)$, and the mobile particle moves in the $z =
0$ plane (Fig.~8).  
From Eq.~(\ref{eq:resist}), the forces on the particles are
\begin{eqnarray}
\left( \begin{array}{c} 
	  \vec{F}^{1} \\
	  \vec{F}^{2} 
          \end{array} \right) 
= - \mu
\left( \begin{array}{cc} 
	  A_{11} & A_{12} \\
	  A_{21} & A_{22} \\
          \end{array} \right)
\left( \begin{array}{c} 
	  \vec{U}^{\infty} - \vec{U}^{1} \\
  	  \vec{U}^{\infty} - \vec{U}^{2} \\
          \end{array} \right) ,
\end{eqnarray}
where $\vec{F}^{i}$ and $\vec{U}^{i}$ are the hydrodynamic force and
velocity acting on particle $i$, and we ignore the $z$ components of
these vectors.  
At these near-touching distance, the contribution from the linear motion of
the particle is much larger ($1/\xi$, where $\xi$ is the gap width)
than that from the rotation ($\ln \xi$), so we ignore the contribution
of rotation of the particle to the forces.
Imposing the force free condition $\vec{F}^{1} = 0$
and the fixing condition $\vec{U}^{2} = 0$,
\begin{eqnarray}
(X_{11}^{A} + X_{12}^{A}) ~ U_{x}^{\infty} & = & X_{11}^{A} ~ U_{x}^{1} 
\nonumber \\
(Y_{11}^{A} + Y_{12}^{A}) ~ U_{y}^{\infty} & = & Y_{11}^{A} ~ U_{y}^{1},
\end{eqnarray}
where $X_{ij}^{A}$ and $Y_{ij}^{A}$ are two particle resistance
functions \cite{kk91}.  When the particles nearly touch,
\begin{equation}
U_{y}^{1} = (1 + {Y_{12} \over Y_{11}}) ~ U_{y}^{\infty} 
          \simeq (1 + {Y_{12} \over Y_{11}}) ~ b,
\end{equation}
where $b$ is the impact parameter as shown in Fig.~8.  
Substituting the near-field form of the resistance matrices \cite{kk91} 
we have 
\begin{equation}
\label{eq:alpha}
U_{1}^{y} \simeq - {3 (A_{11}^{Y}(1) + A_{12}^{Y}(1)) \over \ln \xi} ~ b.
\end{equation}
Here, $A_{ij}^{Y}(1)$ are known constants satisfying $A_{11}^{Y}(1) +
A_{12}^{Y}(1) > 0$, and $\xi$ is the gap distance between the two
particles.  Since $\ln \xi$ is a slowly varying function, we can
ignore its variation.  The above equation becomes,
\begin{equation}
U_{1}^{y} \simeq -\alpha ~ b,
\end{equation} 
where $\alpha$ is a new constant.  Note that the approximations made
are valid at near-touching distance, and within a certain range of
$\xi$.  We can estimate the waiting time as that required for particle
to proceed from $y = y$ to $y = y + \delta y$. Then,
\begin{equation}
\label{eq:tw}
T_w \sim \int_{y}^{y + \delta y} {1 \over U_{1}^{y}} ~ dy 
    \sim {1 \over \alpha} \ln (1 + {\delta y \over y}),
\end{equation}
where we use that $b \simeq y$ near touching distance.  

We determine $\alpha$ and $\delta y$ from the least square fit of the
Eq.~(\ref{eq:tw}) to the measured time for the two particle
simulation.  The waiting time from Eq.~(\ref{eq:tw}) with these
parameters ($\alpha = 0.166$ and $\delta y = 0.329$) is shown along
with measured waiting times in Fig.~7.
The overall shape of the analytic curve agrees very well with the
numerical computations for small values of the impact parameter, while
the deviation at large $d$ is expected, since Eq. (\ref{eq:tw}) holds
only at if the particles nearly touch.

Given the agreement between the behavior of the waiting times in
various systems, it is reasonable to conclude that two-body
interactions control the tail of $T_w$.  Two qualifications are in
order, however.  The porous systems we have considered are relatively
``open'', and it is evident which fixed particle the mobile one
interacts with.  In a densely packed porous medium, particularly one
involving heterogeneous shapes and sizes, some ambiguity may be
present.  Secondly, one may ask about the effects of the other mobile
particles.  When the suspension is dilute, one may distinguish between
the case of a mobile particle approaching another that is waiting in a
pore, the subject of the next section in fact, and the effects of
perturbations in the velocity field induced by more distant suspended
particles.  An accurate treatment of the latter question requires a
much more extensive set of additional simulations than we are able to
provide at this time, but an approximate treatment may be given by
considering the effect of {\em noise} on the waiting time.  We used
the 2d-11 geometry, with $a = 4.5$ and $b = 0.9$, and oscillated the
particle at the lower left to provide a perturbation on the trajectory
of the mobile particle.  The amplitude of the oscillation was $1$ and
its frequency is 0.1, comparable to the time scale of the particle
motion.  The measured waiting time with noise is not substantially
different than without, except that the peak is somewhat rounded, and
we conclude that perturbations of this form do not seem to alter the 
waiting time significantly.

\section{Perturbation of a Waiting Particle}

As discussed in the previous section. a particle may spend significant 
amount of time around a junction, and we now consider the perturbations
induced by other mobile particles in the vicinity, {\em i.e.}, the effect 
of a ``bypassing'' particle on a ``waiting'' particle.  This represents one
special case of the hydrodynamic interaction between two mobile
particles in a porous medium, but a particularly important one in
applications such as deep bed filtration where slow particles are
likely to adhere or be left behind in the filter.

In the previous section, it was shown that the dynamics of particles
moving slowly through pores is controlled by two particle
interactions, so we first consider the effect of a third mobile
particle on a simple two-particle system.  We fix a particle at the
origin, and insert mobile particles at $(-2.2,d_{w},0)$ and
$(-5,d_{p},0)$.  The first mobile particle would wait near the fixed
one if alone, and the second mobile particle perturbs it.  We measure
the waiting time $T_{w}$, again defined as the amount of time the
waiter spends in $-2.1 < x < -1.9$.

In Fig.~9(a), we show the waiting time for $d_{w} = 0.1$, plotted
against $d_{p}$.  For large and small values of $d_{p}$, the waiting
time approaches that of the unperturbed particle ($T_w \sim 10$), but
the notable features in the figure are a minimum at $d_{p} = -0.5$ and
a maximum at $d_{p} = 0.6$.  The origin of these extrema can be
understood by inspecting the trajectories of the particles.  Around
the value of $d_{p}$ at which the minimum of $T_{w}$ occurs, the
perturbing particle passes below the fixed particle and it pushes the
waiting particle {\it away} from the fixed particle.  The waiting
particle then easily escapes from the junction.  Around the local
maximum, the perturbing particle passes above the fixed particle and
pushes the waiting particle {\it towards} it, making it harder for the
waiting particle to escape from the junction.

For smaller values of $d_{w}$, the behavior becomes more
complicated.  We show the waiting time for $d_{w} = 0.01$ against
$d_{p}$ in Fig.~9(b).  Here again, for small and large values of
$d_{p}$, the waiting time approaches that of the unperturbed particle.
However, there are now {\em three} maxima and {\em three} minima in
the waiting time, compared to one each in the previous case.  We label
the minima as A ($d_{p} = -2.0$), B ($d_{p} = -0.5$), and C ($d_{p} =
0.6$), and the maxima as D ($d_{p} = -0.9$), E ($d_{p} = -0.1$), and F
($d_{p} = 2.5$).  The minimum A is the ``push away'' case of $d= 0.1$.
Around minima B and C, the waiting particle, by following the
perturbing particle, escapes from the junction.  The perturbing
particle, which seems to form a temporary bound state with the waiting
particle \cite{bg72}, makes it easier for it to escape.  In other words, it
``leads'' the waiting particle from the junction.  The minimum B (C)
occurs when the perturbing particle leads below (above) the fixed
particle.  We next consider the maxima; at D, the perturbing particle
initially leads the waiting particle.  However, the bond between the
two breaks, when the waiting particle cannot catch up to the leading
particle.  The waiting particle then changes its direction
(``turns''), and proceeds to the opposite channel.  This series of
events increases the waiting time.  The maximum E is due to a ``head
on'' collision.  When all three particles lie close to a straight
line, we expect a large waiting time since any motion orthogonal to
the line, which is essential for escape, will take time.  The last
maximum (F) is due to the ``push toward'' case of $d_{w} = 0.1$.
As $d_{w}$ is further decreased, we observe the same mechanisms
in the qualitative behavior of the waiting time.  We have gone as low as
$d_{w} = 0.0001$, with no new phenomena appearing.

Next we ask whether the above ``moves'' are also observed in more
realistic geometries, by studying the waiting time in the 2d-11
geometry (with $a = 5$ and $b= 0$).  We insert one mobile particle at
$(x_{c}-2.2, y_{c}+d_{w}, 0)$, where $(x_{c}, y_{c})$ are the
coordinates of the center particle.  A mobile perturbing particle is
placed at $(x_{c}-7, y_{c}+d_{p}, 0)$, and we measure the waiting time
of the first mobile particle for a few combinations of $d_{w}$ and
$d_{p}$.  In Fig.~9(c), we plot the waiting time against $d_{p}$ for
$d_{w} = 0.0001$.  The curve is not very different than the previous
case of $d_{w} = 0.01$.  There are three maxima at $d_{p} = -1.4, 0.4$
and $1.6$, and two minima at $d_{p} = -0.1$ and $d_{p} = 0.9$, but the
waiting time does not yet approach to its unperturbed value in the
current range of $d_{p}$.  The origin of the extrema can also be
connected to particular moves of the two particles.  The three maxima
are due to the ``turn'' of the waiting particle.  The motion near the
minimum at $-0.1$ is dominated by the ``push away'' move, and at the
other minimum by the ``lead'' move.  Note that in a this geometry with
many fixed particles, a single trajectory can contain multiple moves.
For example, both the maximum at $0.4$ and the minimum at $0.9$
involve both the lead and turn moves.  Near the maximum, the turn move
gives dominant contribution to the waiting time, while the lead move
dominates near the minimum.

The waiting times for the 2d-11 configuration are not exactly the same
as that for the three particle configuration, not entirely
surprisingly.  Not only is the sequence of moves different, but new
ones are present, and it would be difficult to infer the waiting time
for a realistic porous medium from these simple studies alone.  It is
clear that such three or multi-body interactions have a significant
quantitative effect on transit times for particles through a porous
medium, and this effect could at best be captured in an average way in
simplified models, such as those based on network representations of
porous media.

\section{Relaunching of Trapped Particles}

One of the important mechanisms for the capture of particles suspended
in a fluid flowing through a porous medium (see, e.g.,
\cite{g77,t89}), is geometrical trapping (or straining), where particles
are caught in constrictions smaller than their diameter.  It has been
observed experimentally that such particles can escape (or be
``relaunched'') from the trap, due to another particle passing near
the trapped one \cite{gdhg96a,gdhg96b}.  These authors argue, albeit
without direct evidence, that the relaunching is caused by
hydrodynamic interactions between the two particles.  These
experiments also indicate that relaunching qualitatively changes the
distribution of the trapped particles, and in particular the
efficiency of a filter.  Thus understanding of the relaunching
mechanism and its quantitative measurement are important in
understanding the long time behavior of particulate systems such as
deep bed filters.  We now consider this question within our model
porous media, and show that relaunching does occur with hydrodynamic
interaction only, and then estimate the relaunching rate.

We first study a two dimensional system, where all particles are
confined to the $z = 0$ plane.  We form a trap using two fixed
particles, whose geometry is determined by the distance $r$ between the
two particle centers and $\theta$, the angle between the line
joining the two particle centers and $y$-axis (Fig.~10).  The
coordinates of two particles are $(\pm (r/2) \sin \theta, \pm (r/2)
\cos \theta, 0)$, and the ambient flow is $(1,0,0)$.
We insert a mobile particle in the trap, barely touching the two fixed
particles as shown in Fig.~10.  In the absence of additional particle,
the particle remains in the trap, and its exact coordinate is
determined within the SD simulation.
We then insert another mobile particle at $(-5, d)$, which is carried
near the trap, and ask under what conditions the trapped particle is
dislodged.  Typically, for given values of $r$ and
$\theta$, there is a finite range of $d$ where relaunching is
observed: the $r = 2$ results, for example, are plotted in
Fig.~11. This behavior is qualitatively reasonable, since as $\theta$
increases, the ``barrier'' becomes more aligned with the ambient flow.
(For $r = 2$, if $\theta$ is much larger than $30^\circ$, we
expect the mobile particle does not touch the lower fixed particle,
and moves away from the trap even in the absence of any disturbance.)
We next fix $\theta$ at $10$ degrees, and study the dependence of the
relaunching condition on $r$.  We find that relaunching becomes less
frequent as $r$ increases, and
none at all is observed for $r = 3$, which again can be understood in
terms of the stability of the trapped particle.
In order to calculate the average relaunching probability, one has to
average over all possible trap configurations (over $r$ and $\theta$),
and the trajectory of the perturbing particle (over $d$).  A very rough
estimate of the probability, based on the data similar to that above, is
on the order of a few percent.

An amusing aspect of relaunching is that in the trajectories taken by
the particles it is not the case, as one might expect, that the
perturbing particle ``pushes'' the trapped one away from the trap.
Rather, we find that the perturbing particle always ``leads'' the
trapped particle from the trap (see the previous section for the
precise definition of push and lead).

Next, we consider how relaunching is affected by a porous medium --
the other fixed particles surrounding the trap.  To this end, we
replace the center particle in the 2d-11 geometry by the two-particle
trap just discussed, giving a 12-particle porous medium characterized
by the lattice constant $a$, the vertical shift of the second layer
$b$, the distance between two trap particles $r$, and the tilt angle
$\theta$.  We insert one mobile particle in the trap, and a second
mobile particle at $(-2,d,0)$ just in front of the first layer.  In
general we find that while relaunching still occurs, the effect is
noticeably suppressed by the surrounding particles.  For example, if
we fix $a = 8$, $b = 0$ and $r = 2$ while varying $\theta$ and $d$,
again no relaunching is observed for $\theta = 0$, for $\theta =
10^\circ$ the $d$-interval for relaunching is too narrow to be worth
determining
and for $\theta = 20$ relaunching is observed for the interval $0.5
\le d \le 0.9$, narrower than in the open geometry.  Again the
mechanism at the individual trajectory level is that the perturbing
particle leads the trapped particle away from the trap.  Thus, while
relaunching observed in a model porous medium as well as an open
system, its likelihood is considerably reduced, by a factor of $5 \sim
10$ for these parameters.  In other cases, for example significantly
smaller values of $a$, relaunching is again suppressed if not
eliminated.  In qualitative terms, one might say that the surrounding
fixed particles have the effects of suppressing the velocity
perturbations due to the mobile particle (via porous medium
hydrodynamic screening) as well as constraining the phase space
available for the trapped one.

In order to check whether the dimensionality is relevant to
relaunching, we considered a three dimensional triangular trap,
consisting of three particles, with a mobile particle trapped in the
middle, and the ambient flow field orthogonal to the triangular plane.
We find relaunching occurs in a similar manner, where the trapped
particle follows the ``leading'' particle.  The relaunching rate
differs from the two dimensional cases, however, and given the large
parameter space involved, it is not feasible to provide a quantitative 
estimate of the rate for any given porous medium.

\section{Summary}

We have studied the microscopic behavior of particles suspended in
fluid passing through a porous medium using Stokesian Dynamics
numerical calculations.  The porous medium was constructed by fixing
the positions of certain particles in a fluid, with the appropriate
modifications in the code, and the remainder were allowed to move
under the action of an ambient flow and the hydrodynamic interactions
due to all fixed and moving particles.  We measured the fractional
particle flux for two and three dimensional model porous media.  We
find that the FPF for a channel is, to an excellent approximation,
proportional to its fractional channel width (in two dimensions) or
area (in three). The details of the distribution seem to be highly
dependent on the local geometry, in particular the distribution of
particles at the channel entrances, so it is difficult to draw general
conclusions about its form.

We examined the waiting time distribution for particles moving slowly
in pore junctions, typically caused by a bifurcating flow path due to
a particular fixed particle at the pore boundary.  We compared the
behavior of appropriate particle trajectories in two and three
dimensional model porous media to motion under two-body interactions
alone, treated either numerically or analytically.  The results are
essentially the same in all cases and indicate that the waiting time
is dominated by the interaction between the mobile particle and the
fixed particle which divides the possible flow paths.

We then considered the perturbations to a ``waiting'' particle near a
junction due to a second mobile ``bypassing'' particle.  We find that
the interaction between the two particles leads to an unexpected and
rich behavior, displaying several types of motions as the two mobile
particles interact with each other.  An important related quantity is
the relaunching rate of geometrically trapped particles, representing
the likelihood that a particle caught in a geometrical trap is
released by the effects of a second particle nearby.  We constructed
simple traps in two and three dimensions, in which particles are
trapped in a constriction.  We find that the trapped particles can
indeed be relaunched from the trap by this mechanism, and that the
relaunching rate depends sensitively on the surrounding geometry of
the trap.

A principal motivation for this work is to develop computational
approaches to the dynamics of processes such as deep bed filtration,
where there are innumerable microscopic details, by replacing them by
a coarser-grained but tractable network description.  Network modeling
of fluid or passive tracer flow in porous media seems to capture many
of the relevant aspects \cite{ba88,sah93} and we have been guided by
the detailed experimental observations cited above in identifying pore
scale mechanisms to understand.  The results in this paper have
succeeded in part, in that we have shown that effects such as
hydrodynamic relaunching have a firm basis, and explored many of the
microscopic interactions controlling filtration dynamics.  We are
limited principally by the range of geometries we can construct, in
particular the restriction to monodisperse particles, and by the
relative slowness of SD computations in general.  Future work will, we
trust, alleviate these restrictions.

\section*{Acknowledgment}

We thank J. Brady and T. Phung for providing the Stokesian Dynamics
code, and for helping us with its use, and E. Guazzelli for extensive 
discussions of the filtration experiments which motivated this study.
This work was supported by the Department of
Energy under Grant No. DE-FG02-93-ER14327.  One of us (J.L.) is
supported in part by SNU-CTP and Korea Science and Engineering
Foundation through the Brain-Pool program.  The simulations were
performed at the Systems Engineering Research Institute in Korea.

\newpage

\section*{Figure Captions}

\begin{description}

\vfill
\centerline{\psfig{figure=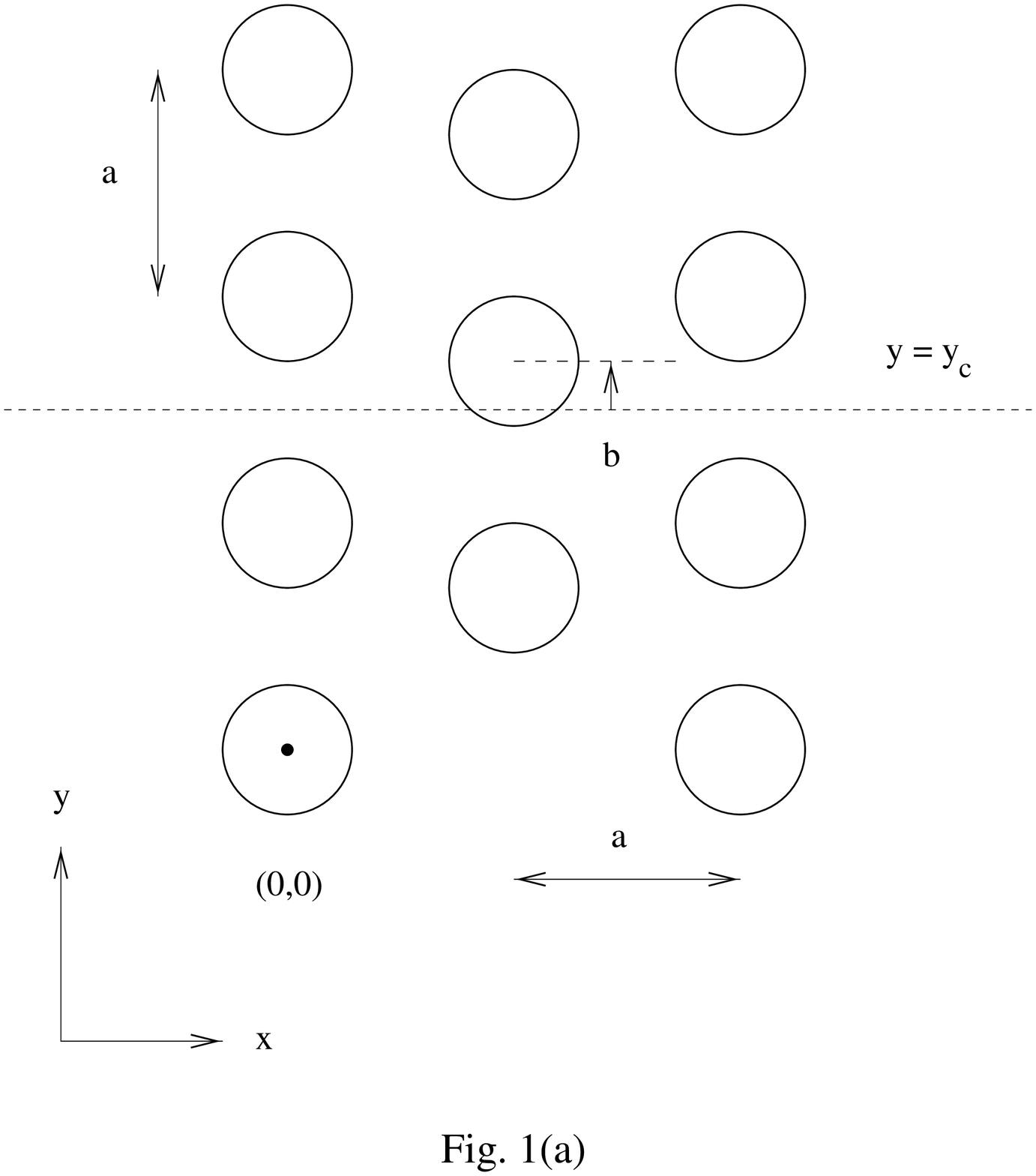,angle=0,width=5in}}
\centerline{\psfig{figure=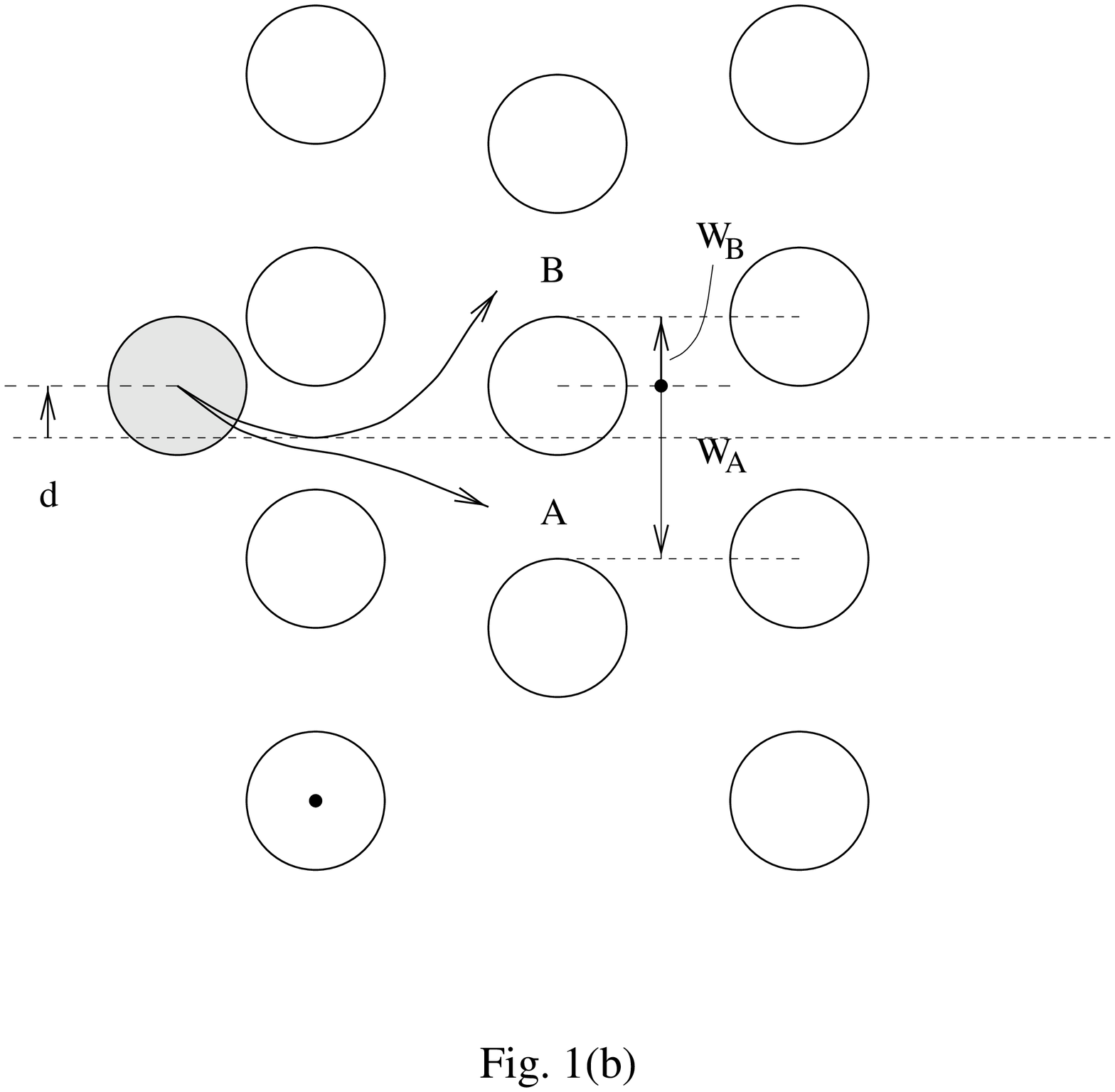,angle=0,width=5in}}

\item [Fig.~1:] (a) A two dimensional model porous medium
consisting of $11$ spheres in a plane,  characterized by the ``lattice
constant'' $a$ and the vertical displacement $b$.  (b) Possible trajectories
A and B of a moving particle (shaded), initially at $(-2,y_{c}+d,0)$, 
in the ambient velocity field $(1,0,0)$.
\vfill
\newpage

\vfill
\centerline{\psfig{figure=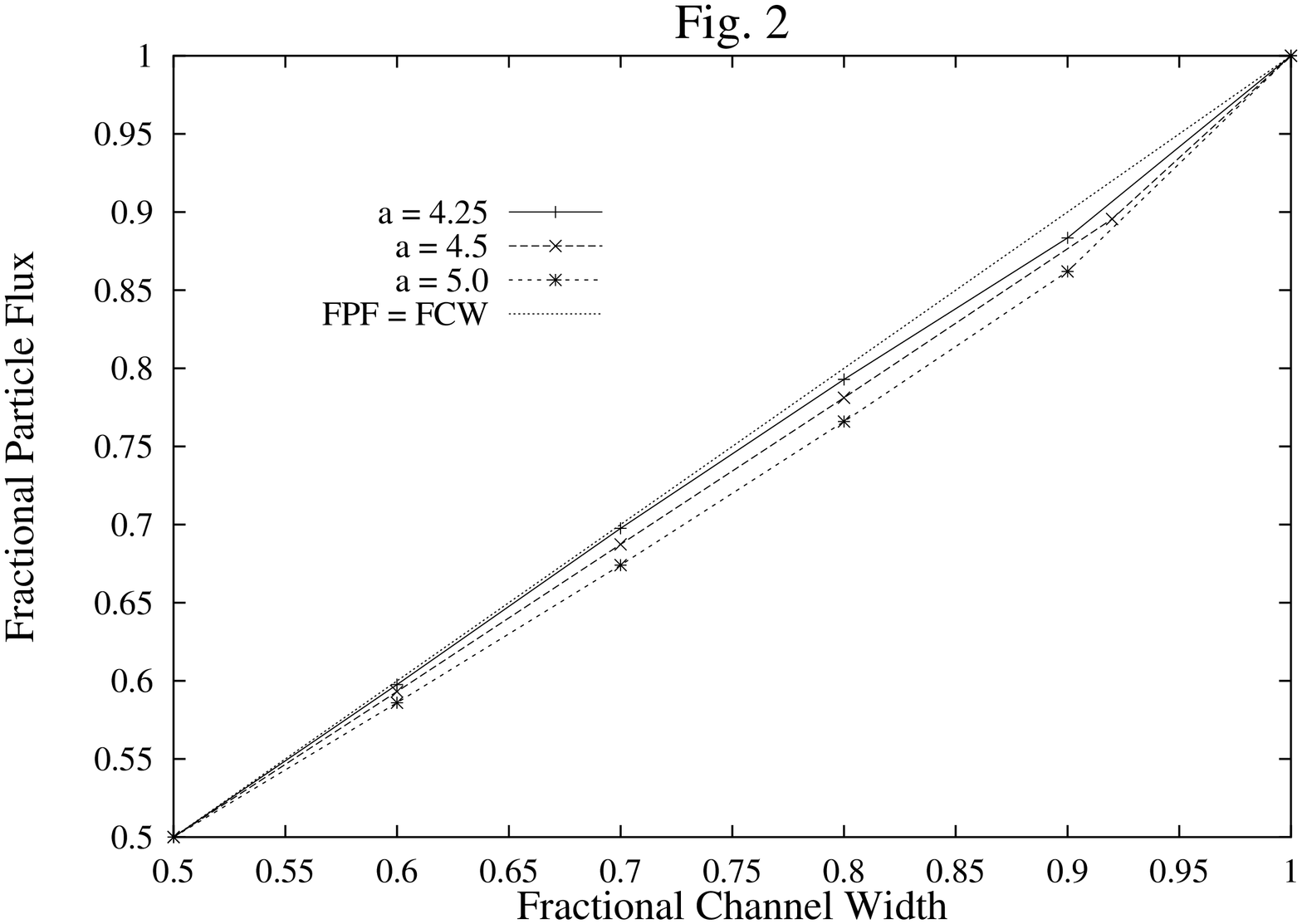,angle=270,width=5in}}

\item [Fig.~2:] The fractional particle flux (FPF) for channel A
plotted against its fractional channel width (FCW) for the $2d-11$ geometry,
for $a = 4.25, 4.5$ and $5.0$.  The dotted line represents FPF = FCW.
\vfill
\newpage

\vfill
\centerline{\psfig{figure=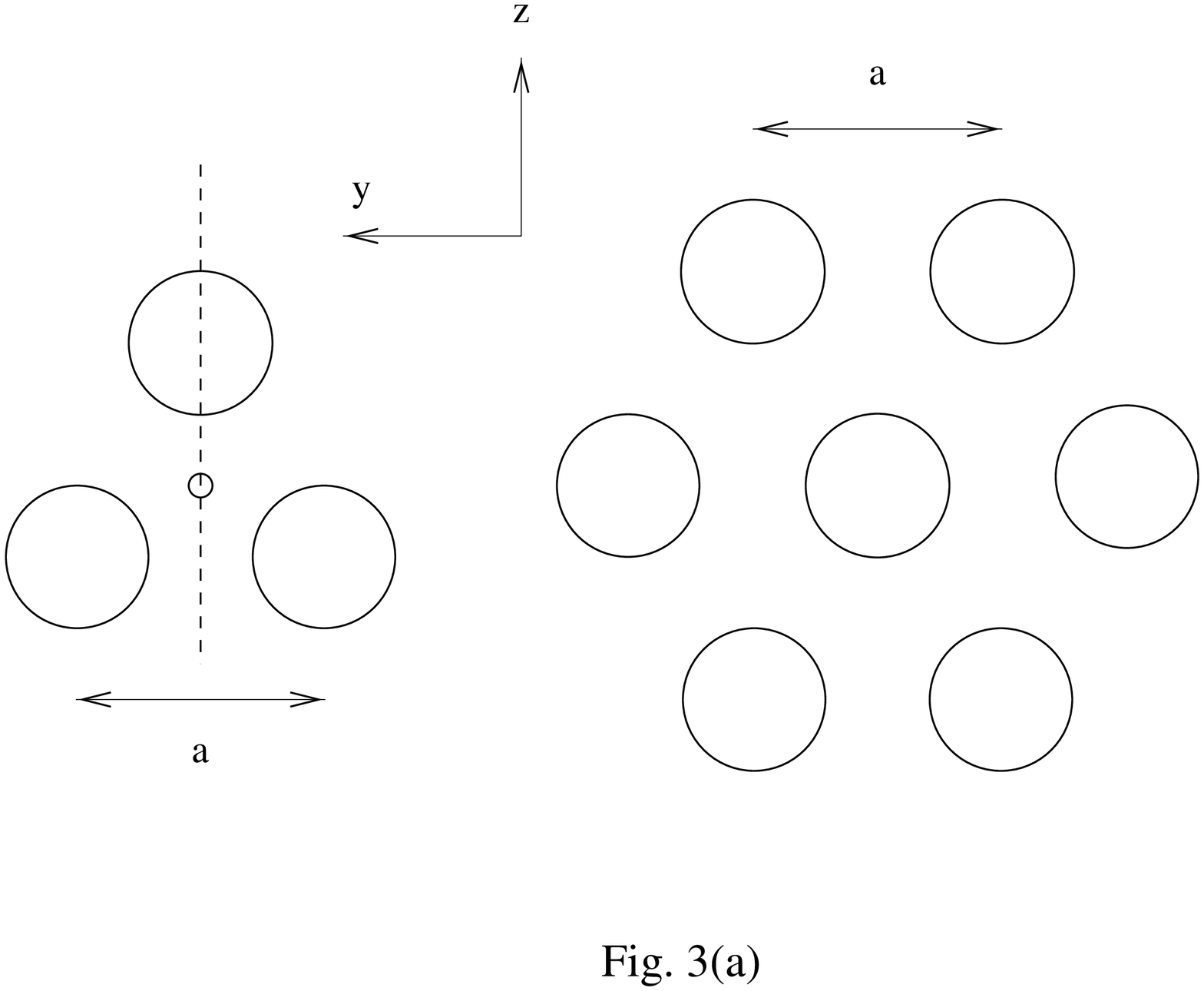,angle=270,width=5in}}
\centerline{\psfig{figure=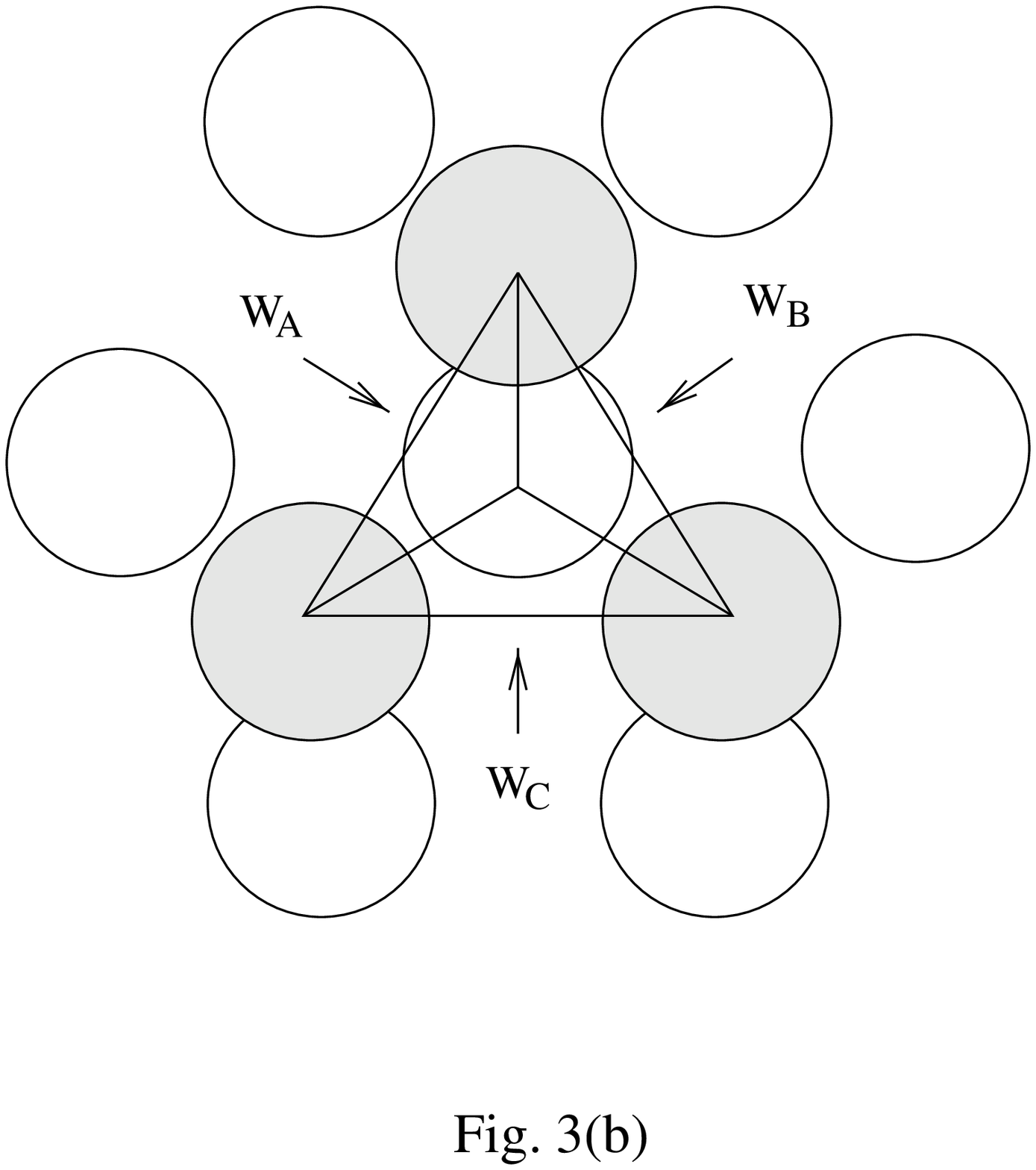,angle=270,width=5in}}

\item [Fig.~3:] (a) The first and second layers, respectively, of a three
dimensional porous medium. The third layer is identical to the first, and they
are spaced in the (out-of-plane) $x$ direction by $a\sqrt{2/3}$,  
where the (in-plane) lattice constant is $a$.  (b) A mobile
particle is initially placed within the triangular ``unit cell'' at a small
value of $x$, and then passes through one
of the three channels (A, B or C).  The figure here shows the projection onto
the $y-z$ plane.  
\vfill
\newpage

\vfill
\centerline{\psfig{figure=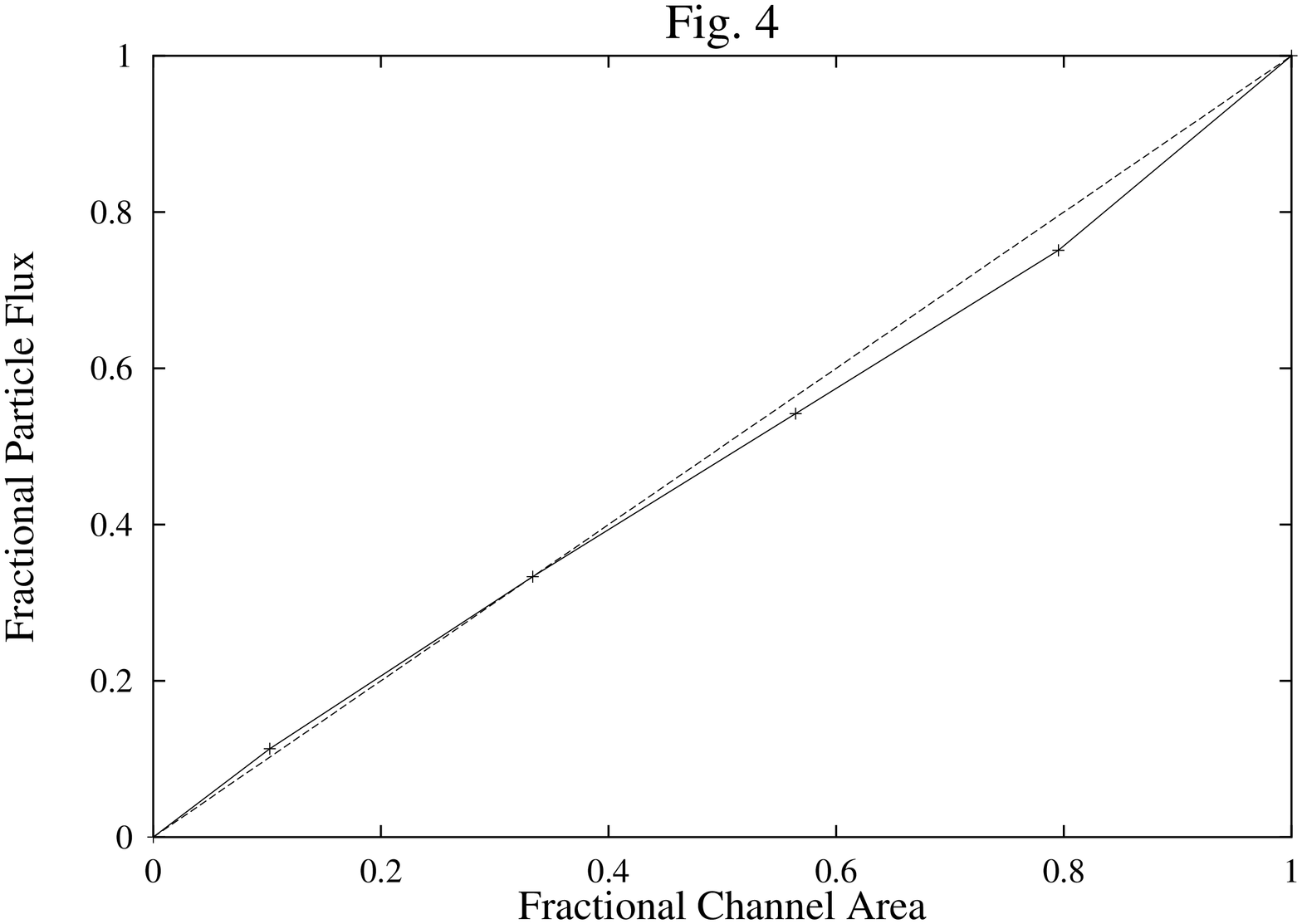,angle=270,width=5in}}

\item [Fig.~4:] Fractional particle flux (FPF) passing through
channel C against its fractional channel area (FCA) for the $3d-11$
geometry with $a = 10$.  The dotted line represents FPF = FCA.
\vfill
\newpage

\vfill
\centerline{\psfig{figure=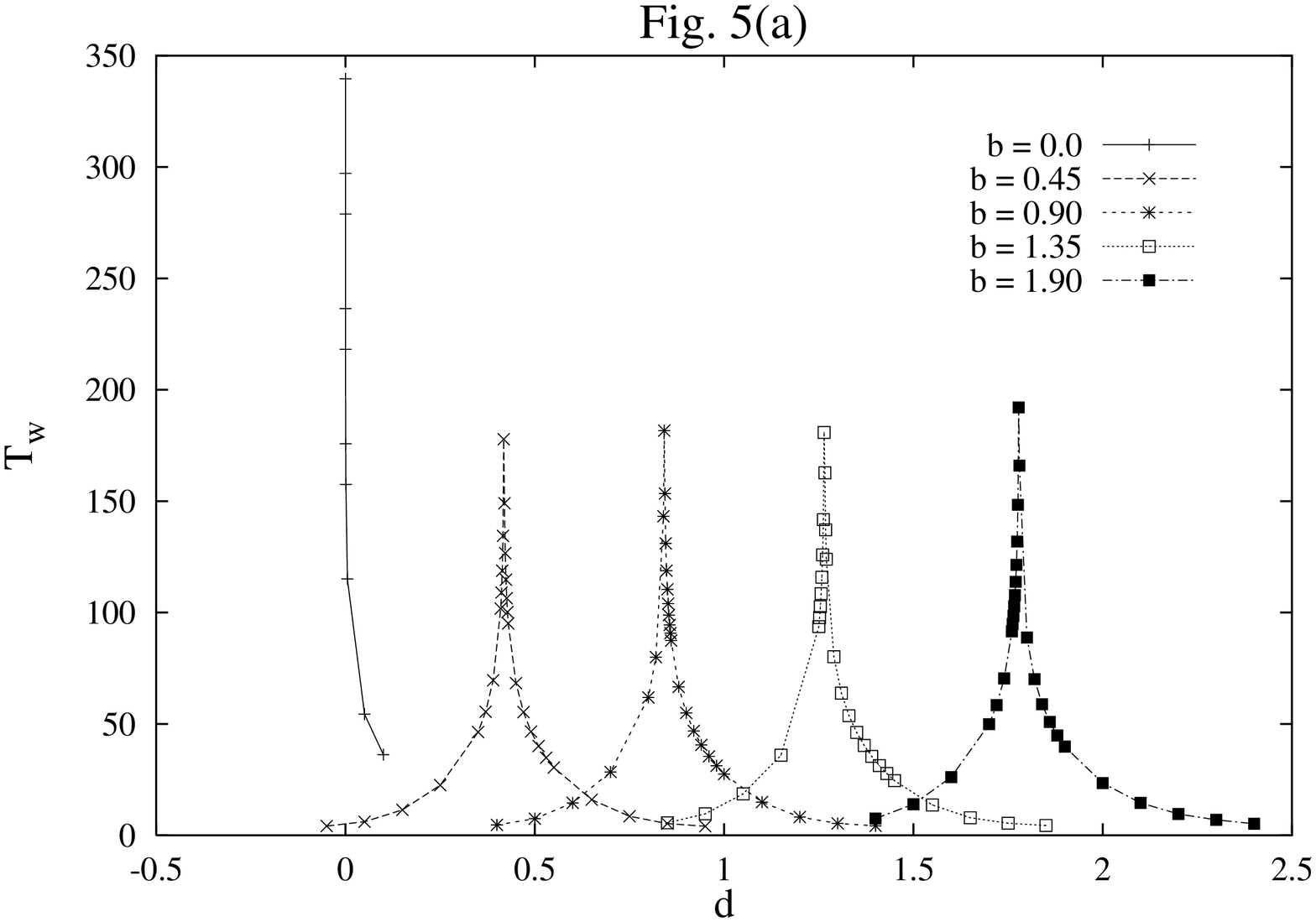,angle=270,width=5in}}
\centerline{\psfig{figure=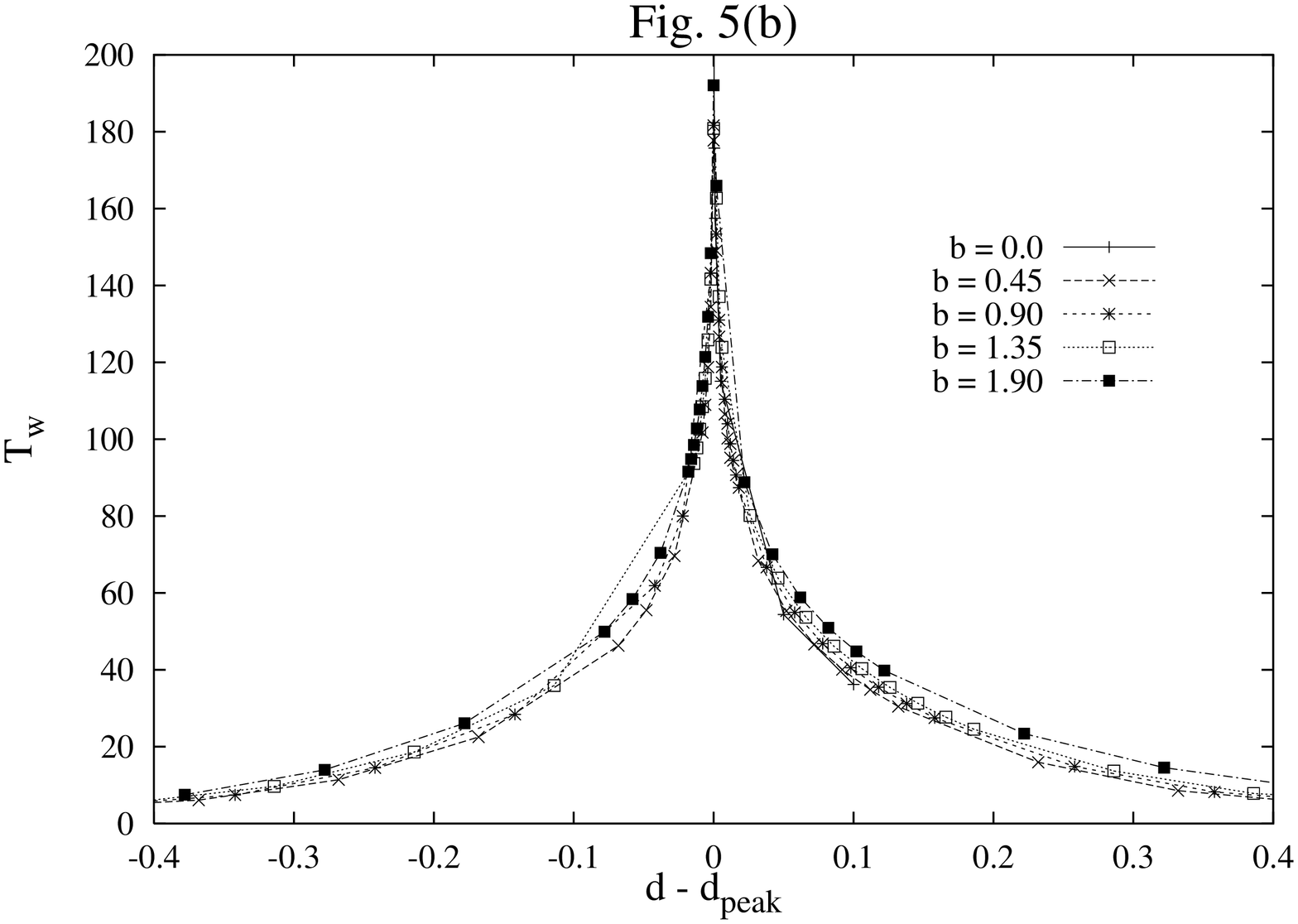,angle=270,width=5in}}

\item [Fig.~5:] (a) Waiting time $T_{w}$ for the 2d-11 geometry
with $a = 4.5$ and several values of $b$.  (b) The curves in (a) are
shifted by $d_{\rm peak}$, whereupon they approximately collapse into a
single master curve.
\vfill
\newpage

\vfill
\centerline{\psfig{figure=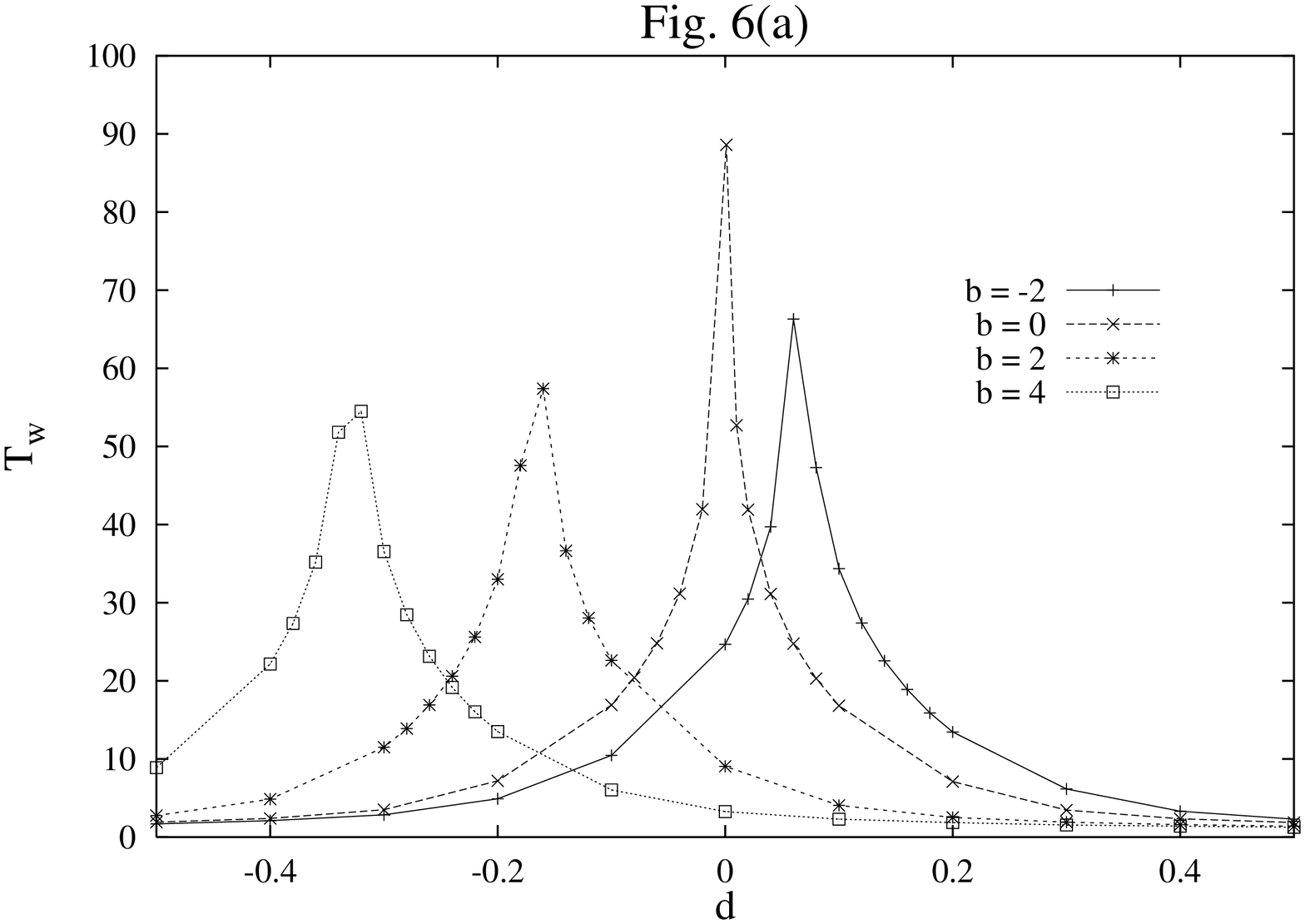,angle=270,width=5in}}
\centerline{\psfig{figure=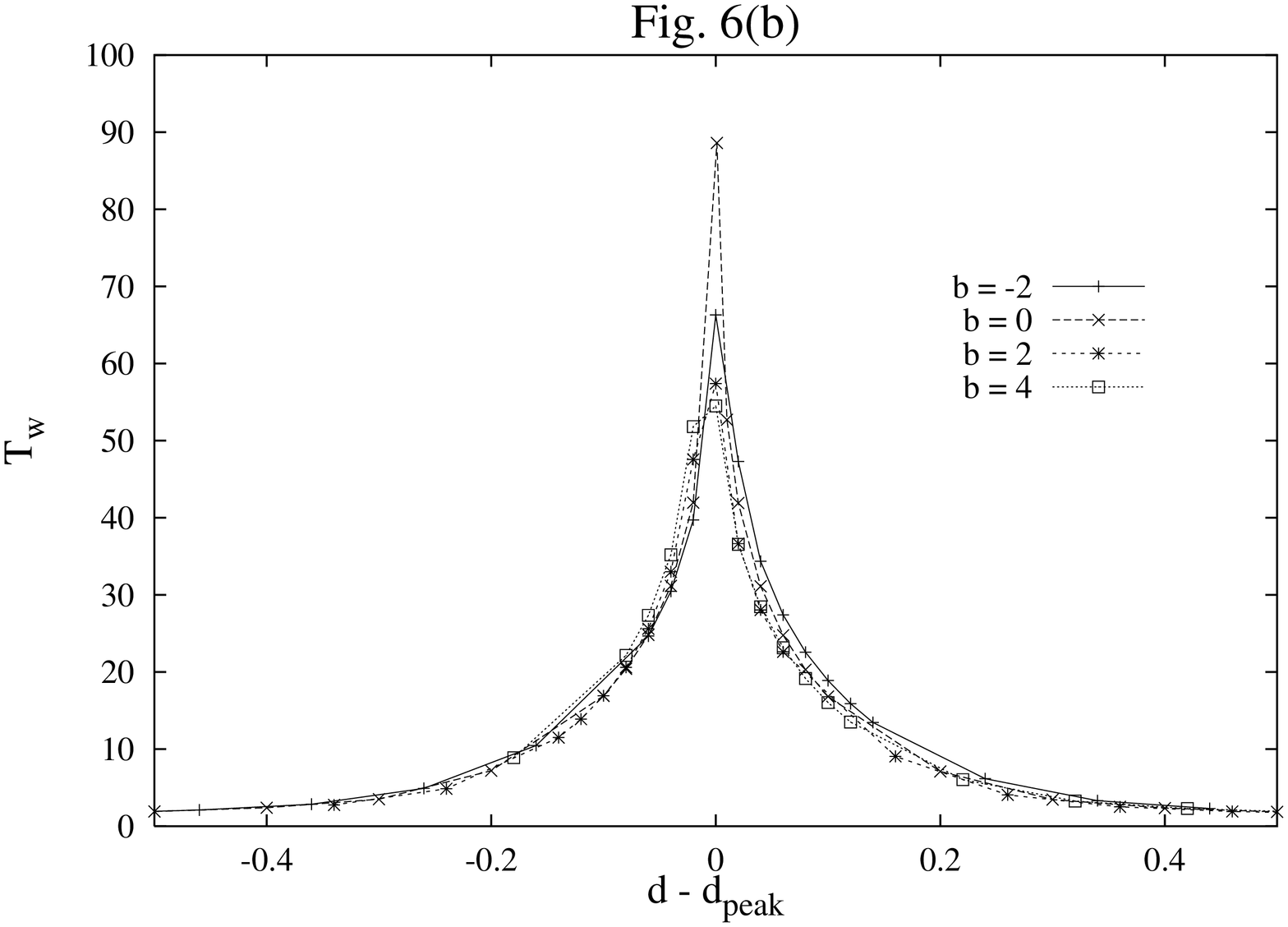,angle=270,width=5in}}

\item [Fig.~6:] (a) Waiting time $T_{w}$ for the 3d-13 geometry plotted
against $d$, with $a = 10$ and four values of $b$.  (b) The curves in
(a) are shifted by $d_{\rm peak}$, and again seem to collapse to
a single master curve.
\vfill
\newpage

\vfill
\centerline{\psfig{figure=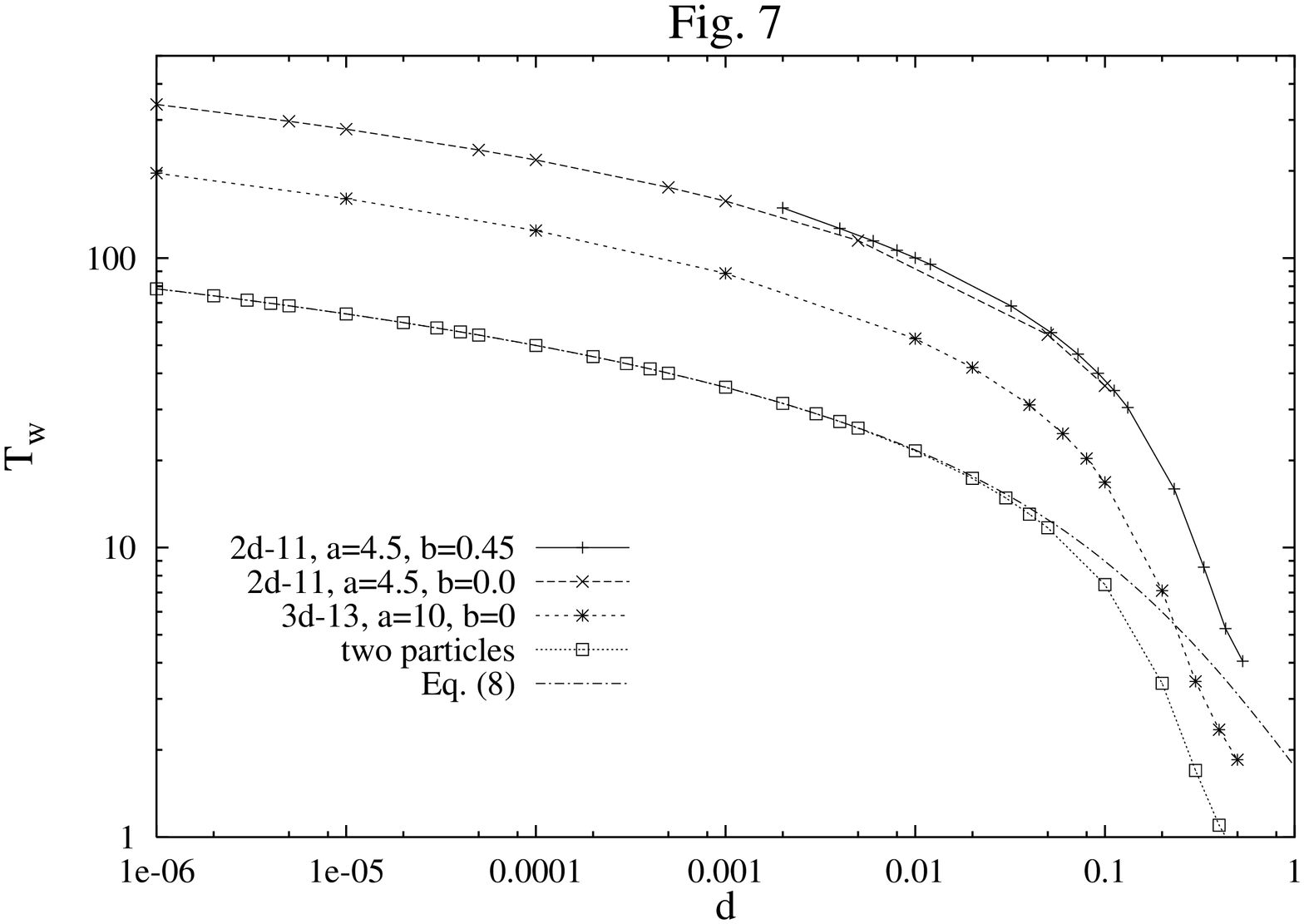,angle=270,width=5in}}

\item [Fig.~7:] Waiting time for a two particle system is shown
together with those for the 2d-11 ($a = 4.5, b = 0$ and $b = 0.45$)
and the 3d-13 ($a = 10, b = 0$) geometries, along with the
analytic estimate Eq. (\ref{eq:tw}) with $\alpha = 0.166$ and $\delta
y = 0.329$.
\vfill
\newpage

\vfill
\centerline{\psfig{figure=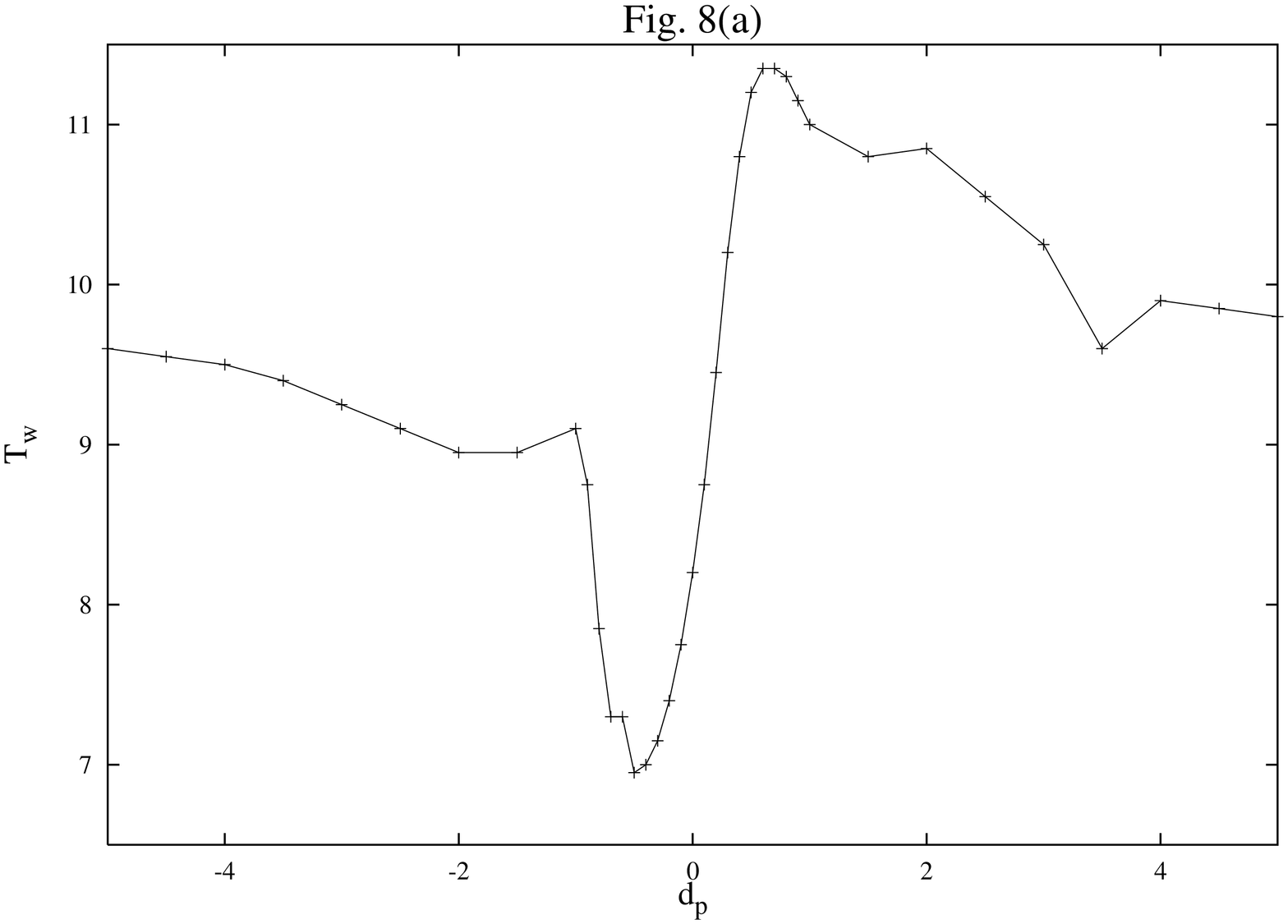,angle=270,width=5in}}
\centerline{\psfig{figure=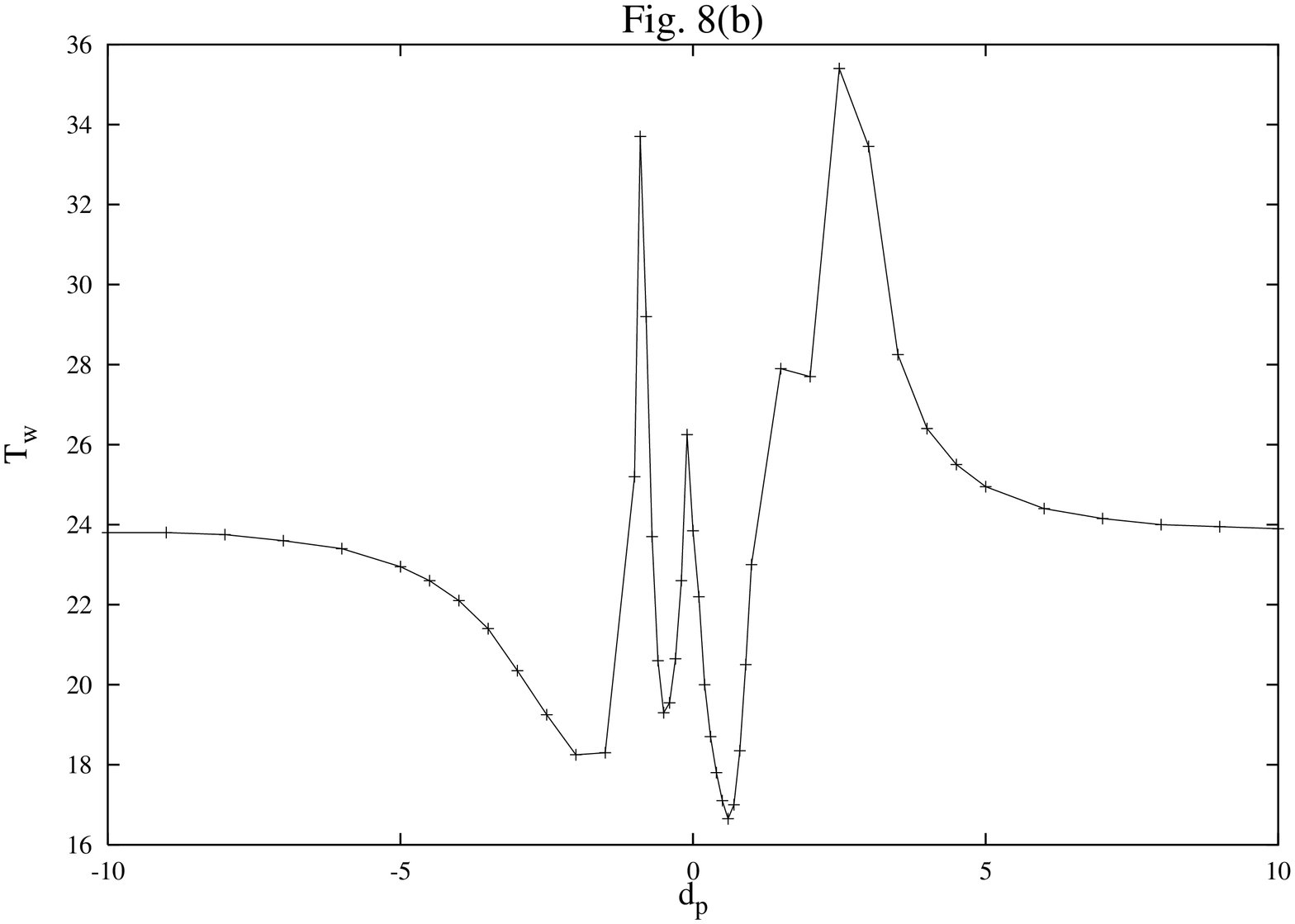,angle=270,width=5in}}
\centerline{\psfig{figure=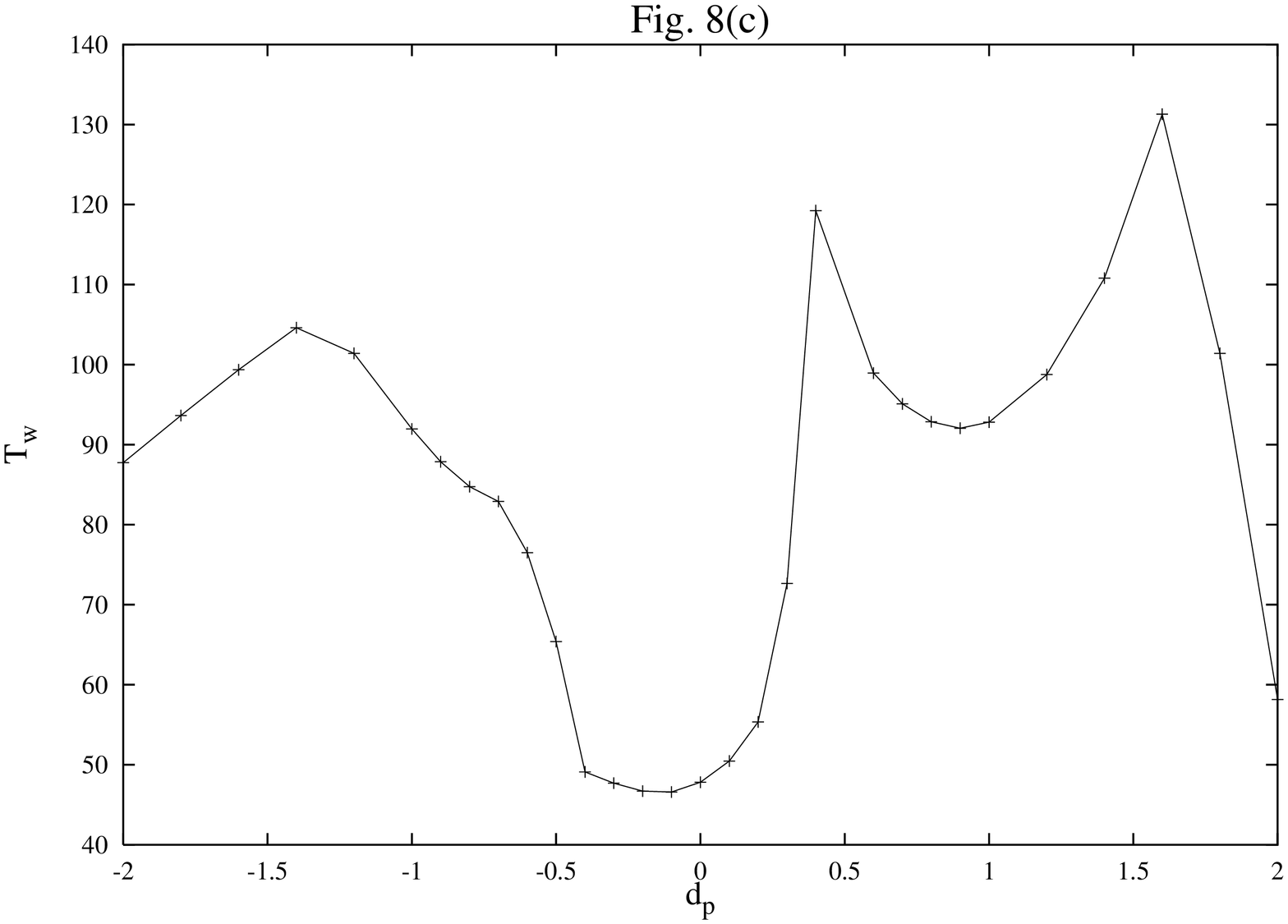,angle=270,width=5in}}

\item [Fig.~8:] Geometry of the two particle system used for the analytical 
estimate of $T_{w}$.  A particle (shaded) is fixed at $(0,0)$, and a mobile
particle starts $d$ away from it,  in the ambient velocity field 
$U^{\infty} = (\cos \theta, \sin \theta, 0)$.
\vfill
\newpage

\vfill
\centerline{\psfig{figure=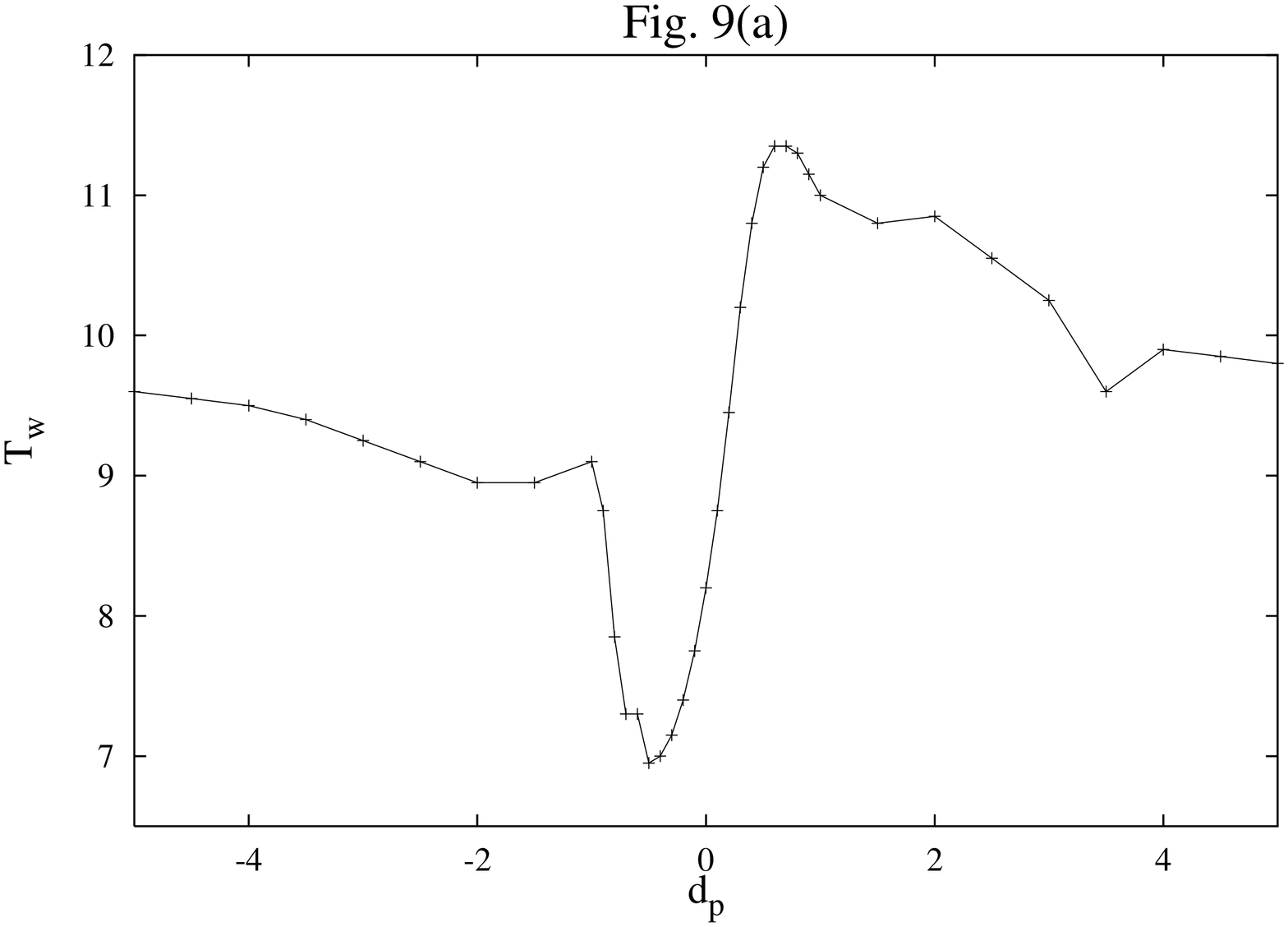,angle=270,width=5in}}
\centerline{\psfig{figure=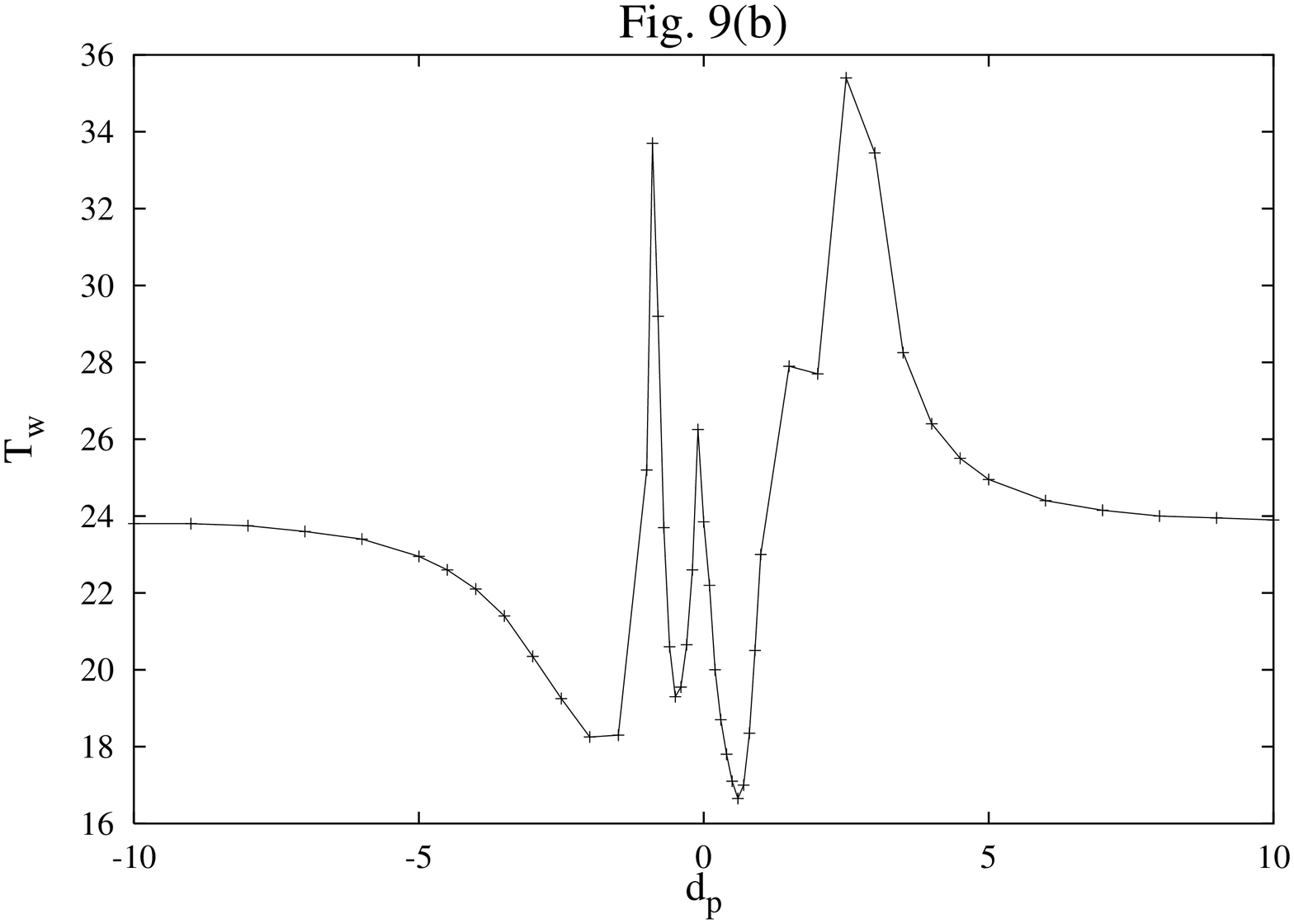,angle=270,width=5in}}
\centerline{\psfig{figure=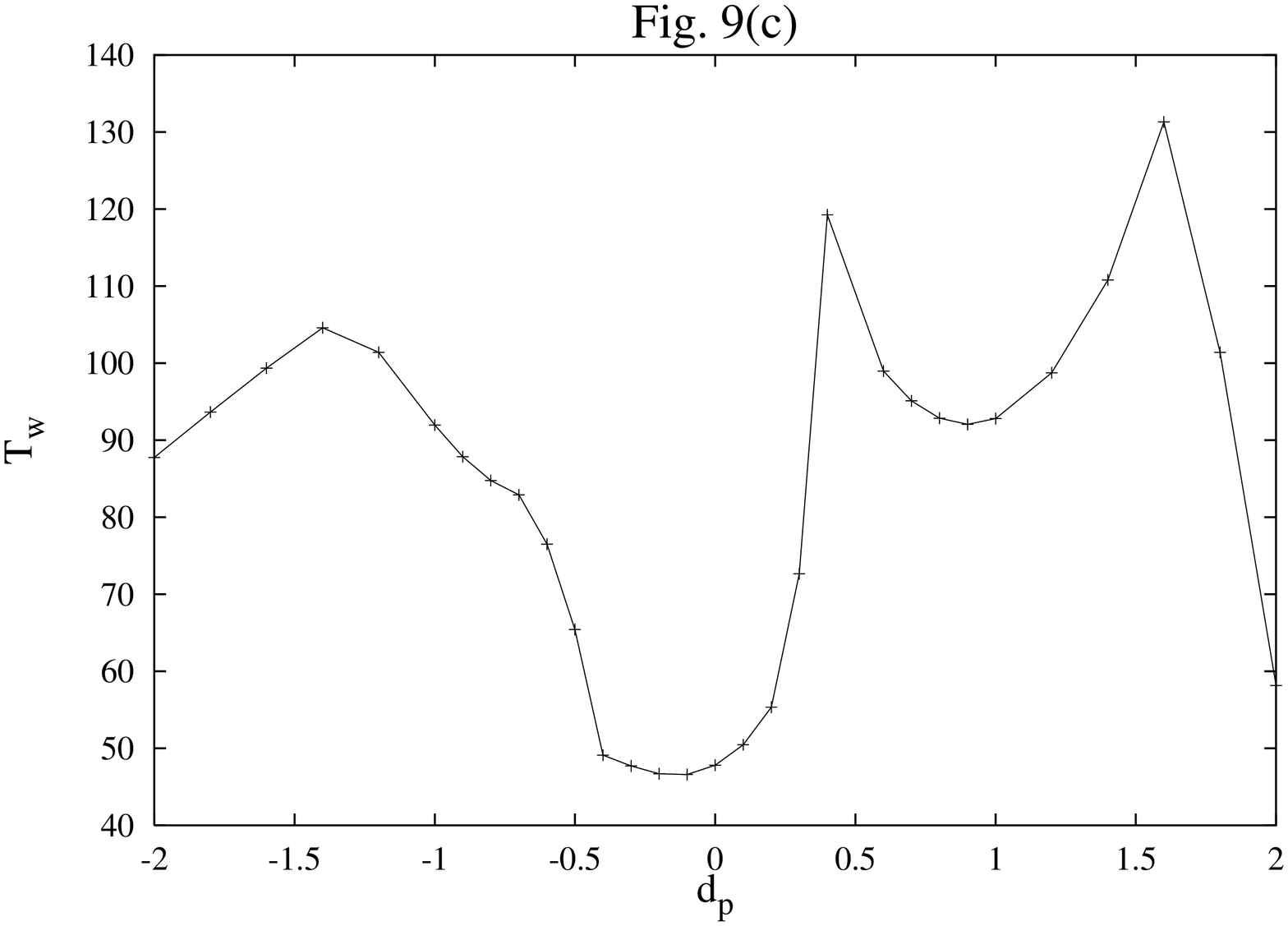,angle=270,width=5in}}

\item [Fig.~9:] Waiting time distributions for the perturbation of a 
slow particle.  A mobile particle is at $(-2.2,d_{w},0)$ in front of
a fixed particle, while a second mobile particle starting at
$(-5,d_{p},0)$ passes by.  The waiting time for the first
particle is plotted against $d_{p}$ for (a) $d_{w} = 0.1$, and (b) $d_{w}
= 0.01$.  (c) The waiting time for a mobile particle in the 2d-11
geometry, in the presence of another mobile particle.  The variable
$d_{p}$ is roughly the impact parameter of the second mobile particle.
\vfill
\newpage

\vfill
\centerline{\psfig{figure=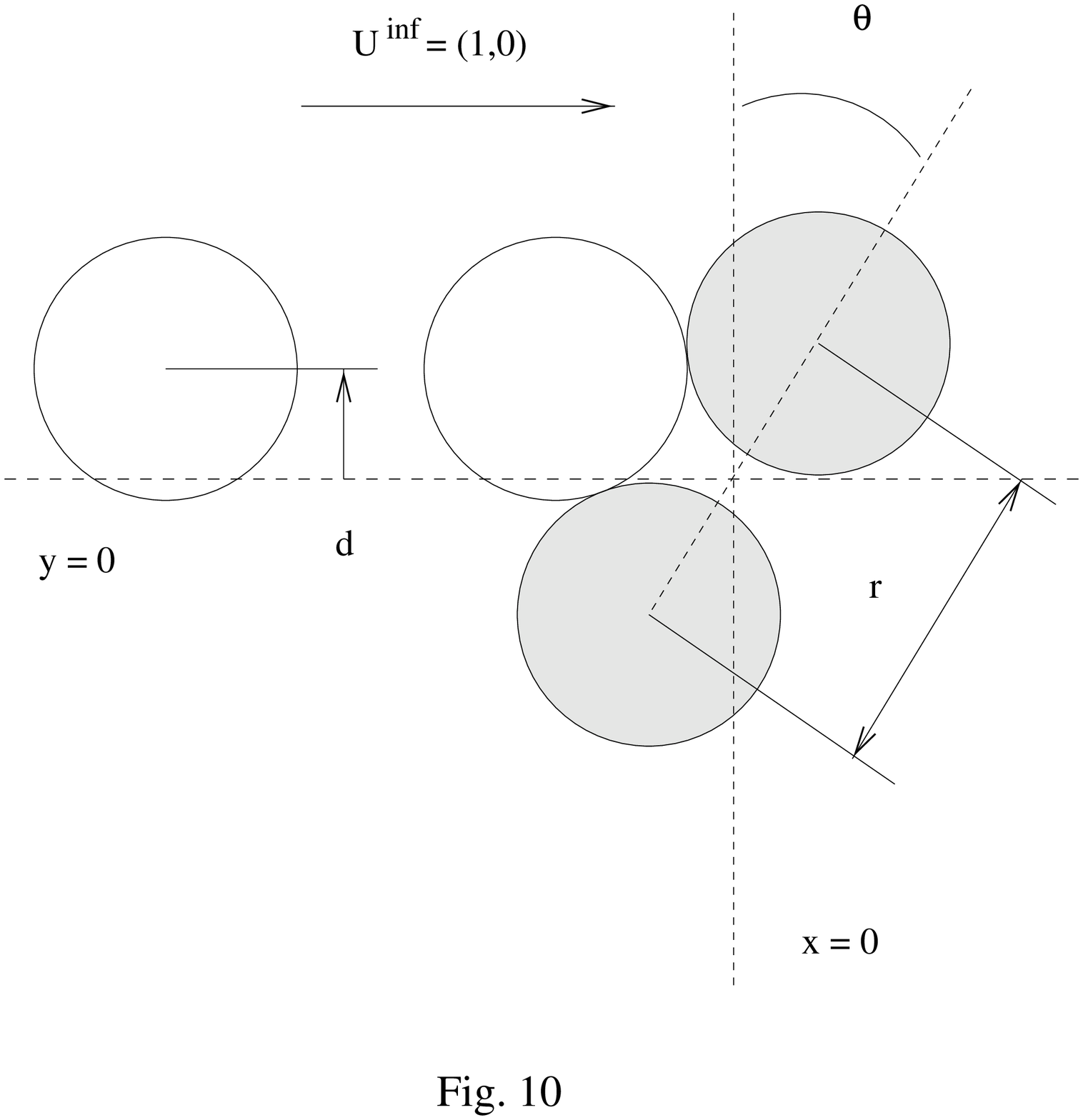,angle=0,width=5in}}

\item [Fig.~10:] Geometry for studying the relaunching of trapped
particles.  A mobile particle is in a model two-particle trap
(marked with grey shade) characterized by $r$ and $\theta$, while another
mobile particle starting at $(-5,d,0)$ passes near the trap.
\vfill
\newpage

\vfill
\centerline{\psfig{figure=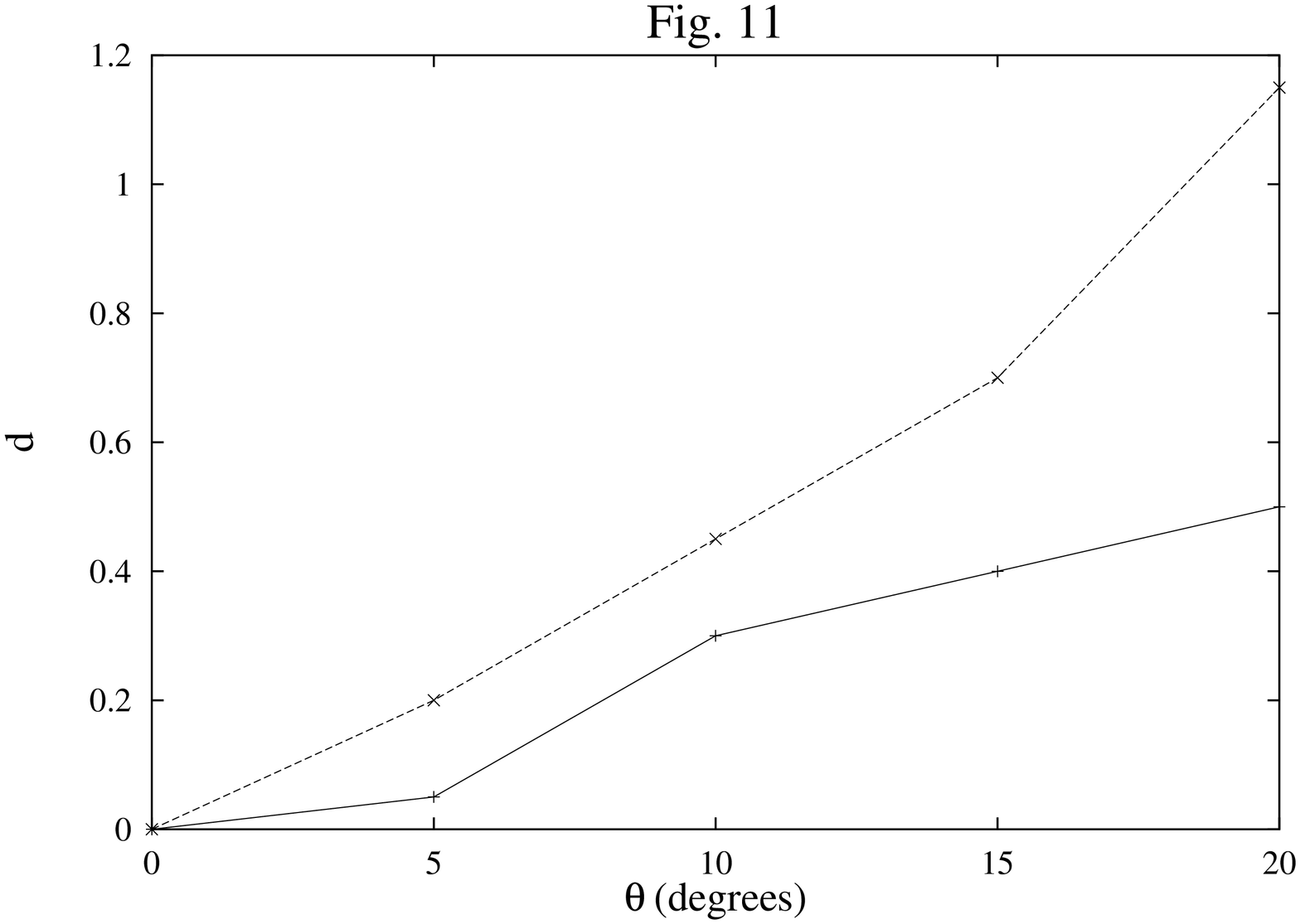,angle=270,width=5in}}

\item [Fig.~11:] The range of $d$ for which relaunching occurs plotted
against $\theta$, for the geometry in Fig.~10 with $r = 2$.  The
relaunching occurs in the region bounded by the two lines.
\vfill
\newpage

\end{description}


\begin{thebibliography}{99}
\bibitem{note} The assumption of the uniform distribution is, in
general, not correct.  We expect that, for a given geometry, the
distribution can be measured by following many trajectories for large
period of time.  However, the distribution we measure by the method is
too sensitive to small changes in the geometry, and it is not clear
that such distribution can serve as a better approximation.
\end{thebibliography}
\end{document}